\documentclass []{aa}

\usepackage{psfig}
\usepackage{times}
\usepackage{graphics}
\usepackage{epsf}

\begin{document}


\title{Morphology and kinematics of Lynds 1642
\thanks{Based on observations collected at the European Southern Observatory}
}

\subtitle{Multivariate analysis of CO maps of a translucent cloud}

\author{
D.~Russeil \inst{1,2}
\and M.~Juvela \inst{1}
\and K.~Lehtinen \inst{1}
\and K.~Mattila \inst{1}
\and P.~Paatero \inst{3}
}

\offprints{D.~Russeil}

\institute{
$^1$Helsinki University Observatory, T\"ahtitorninm\"aki, P.O.Box 14,
FIN-00014 University of Helsinki, Finland \\
$^2$Laboratoire d'Astrophysique de Marseille, 2 Place Le Verrier, 13004 Marseille, France\\
$^3$University of Helsinki, Department of Physical Sciences, P.O.Box 64,
FIN-00014 University of Helsinki, Finland
}

\titlerunning{Kinematics and morphology of L1642}

\date{Received........1999; accepted.......1999}

\abstract{ The high latitude translucent molecular cloud L1642 has been mapped
in the J=1-0 and J=2-1 transitions of $^{12}$CO, $^{13}$CO and C$^{18}$O using
the SEST radio telescope. We have analysed the morphology and velocity
structure of the cloud using the Positive Matrix Factorization (PMF) method.
The results show that L1642 is composed of a main structure at radial velocity
0.2\,km\,s$^{-1}$ while the higher velocity components at $\sim$0.5 and
1.0\,km\,s$^{-1}$ form an incomplete ring around it, suggesting an expanding
shell structure.
Fainter emission extends to the north with a still higher velocity of up to
1.6\,km\,s$^{-1}$. Such a velocity structure suggests an elongated morphology
in the line of sight direction. The physical properties of the cloud have been
investigated assuming LTE conditions, but non-LTE radiative transfer models
are also constructed for the $^{13}$CO observations. We confirm that L1642
follows an $r^{-1}$ density distribution in its outer parts while the
distribution is considerably flatter in the core. The cloud is close to virial
equilibrium. \\
In an Appendix the PMF results are compared with the view obtained through the
analysis of channel maps and by the use of Principal Component Analysis (PCA).
Both PMF and PCA present the observations as a linear combination of
basic spectral shapes that are extracted from the data. Comparison of the
methods shows that the PMF method in particular is able to produce a presentation
of the complex velocity that is both compact and easily interpreted.
 }

\maketitle

\section{Introduction}

High latitude translucent molecular clouds are ideal targets for the study of
the interactions between different phases of the interstellar medium which are
subjected to the general interstellar radiation field. The visual extinction
of these clouds is modest ($A_{V}\sim$1-3\,mag) which places them between dark
and diffuse clouds. As they are minimally affected by embedded star-formation
the external radiation field is the dominant heating source. Furthermore, due
to the high galactic latitude of the clouds, observations are seldom
contaminated by foreground or background emission.  The cloud L1642 (=MBM20)
is a prototype of this class of clouds. It is located at galactic coordinates
$l$=210.9$\degr$, $b$=-36.5$\degr$. The distance determination by Hearty et
al. (\cite{hearty00}) places it between 112 and 160\,pc.  Based on X-ray
measurements, Kuntz et al. (\cite{kuntz97}) suggested that L1642 is within or
close to the edge of the local bubble, which in this direction is at about
140\,pc according to Sfeir et al. (\cite{sfeir99}). This is in good agreement
with the direct measurement and we adopt 140\,pc for the distance of L1642.
Based on previous multiwavelength data (Liljestr\"om and Mattila
\cite{liljestrom88}; Taylor et al. \cite{taylor82}; Laureijs et al.
\cite{laureijs87}; Liljestr\"om \cite{liljestrom91}) L1642 appears as a cool
and quiescent cloud. It is associated with a much larger HI cloud (over
4$\degr$) with cometary structure. This is also seen in the IRAS 100\,$\mu$m
image. The tail is perpendicular to the Galactic plane, and points towards
the plane.

The IRAS point sources IRAS~04325-1419 and IRAS~04237-1419 make L1642 one of the
closest star forming clouds. The source colours suggest that they are T-Tauri
stars. In particular, IRAS~04237-1419 shows Herbig-Haro characteristics
(Sandell et al. \cite{sandell87}; Reipurth \& Heathcote \cite{reipurth90}).
However, the low luminosity of the source ($L<$0.6 L$_{\odot }$) means that it will have
only a very local influence.

In this paper, we present an analysis of L1642, making use of new CO
observations with a wider map area and/or denser grid than in the previous
studies. After the description of observations in Sect.~\ref{sect:observations},
Sect.~\ref{sect:morphology} is devoted to the analysis of morphology and
kinematics using the Positive Matrix Factorization method.
The analysis is also carried out using another multivariate method, the
Principal Component Analysis (Murtagh \& Heck \cite{MH87}) and by channel
maps. A comparison of the analysis methods and some comments on their relative
strengths and weaknesses are given in an Appendix.
In Sect.~\ref{sect:correlations} we discuss the correlations between the
observed lines, and in Sect.~\ref{sect:colden} the physical properties of the
cloud are derived. Finally, in Sect.~\ref{sect:conclusions} we present our
conclusions.

\section{Observations and data reduction} \label{sect:observations}

Molecular line observations were made by K.L. in August 1999. The transitions
J=1--0 and J=2--1 of $^{12}$CO, $^{13}$CO and C$^{18}$O were observed with the
Swedish-ESO Submillimetre Telescope (SEST). The receivers were cooled Schottky
diode mixers operated in 80-116\,GHz and 216-245\,GHz bands. The channel
separation of the AOS spectrometer was 0.11\,km\,s$^{-1}$ for the $J$=1--0 
transition and
0.06\,km\,s$^{-1}$ for $J$=2--1 transition (Booth et al. \cite{booth89}).
Observations were made in frequency switching mode. The calibration was done
using the chopper-wheel method (Ulich \& Haas \cite{ulich76}) and checked by
observing the Orion KL nebula. Antenna temperatures ($T_{\rm A}^{\ast }$) were
converted to radiation temperatures ($T_{\rm R}^{\ast }$) using the Moon
efficiency.  The half power beam widths are 45$\arcsec$ and 23$\arcsec$ for
the 1-0 and 2-1 transitions, respectively.  The grid spacing was 3$\arcmin$.
The grid positions were planned to be identical to the map positions of our
200\,$\mu$m ISOPHOT map (Lehtinen et al., in preparation).

\begin{figure}
\psfig{file=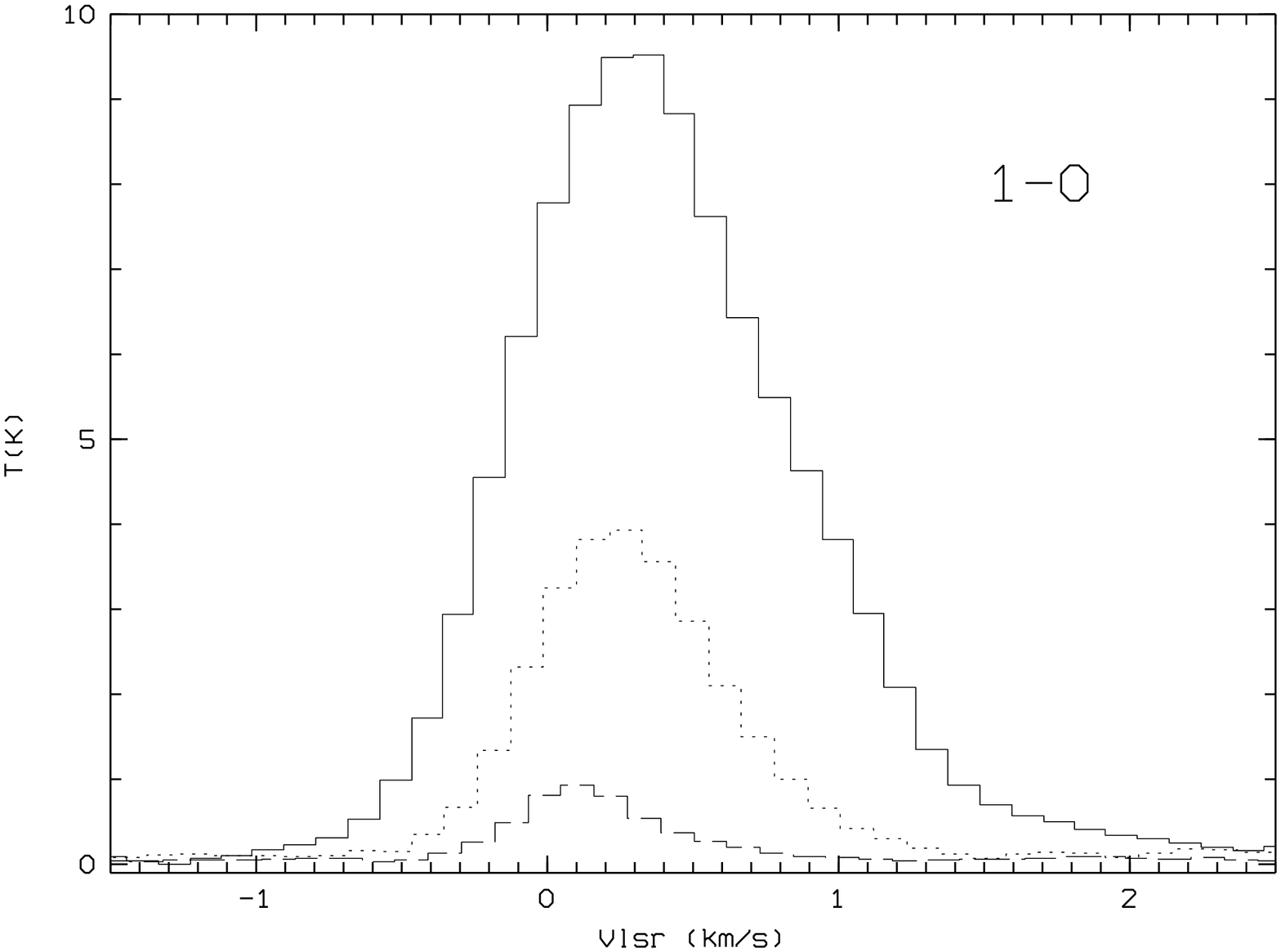,width=8cm,angle=270}
\psfig{file=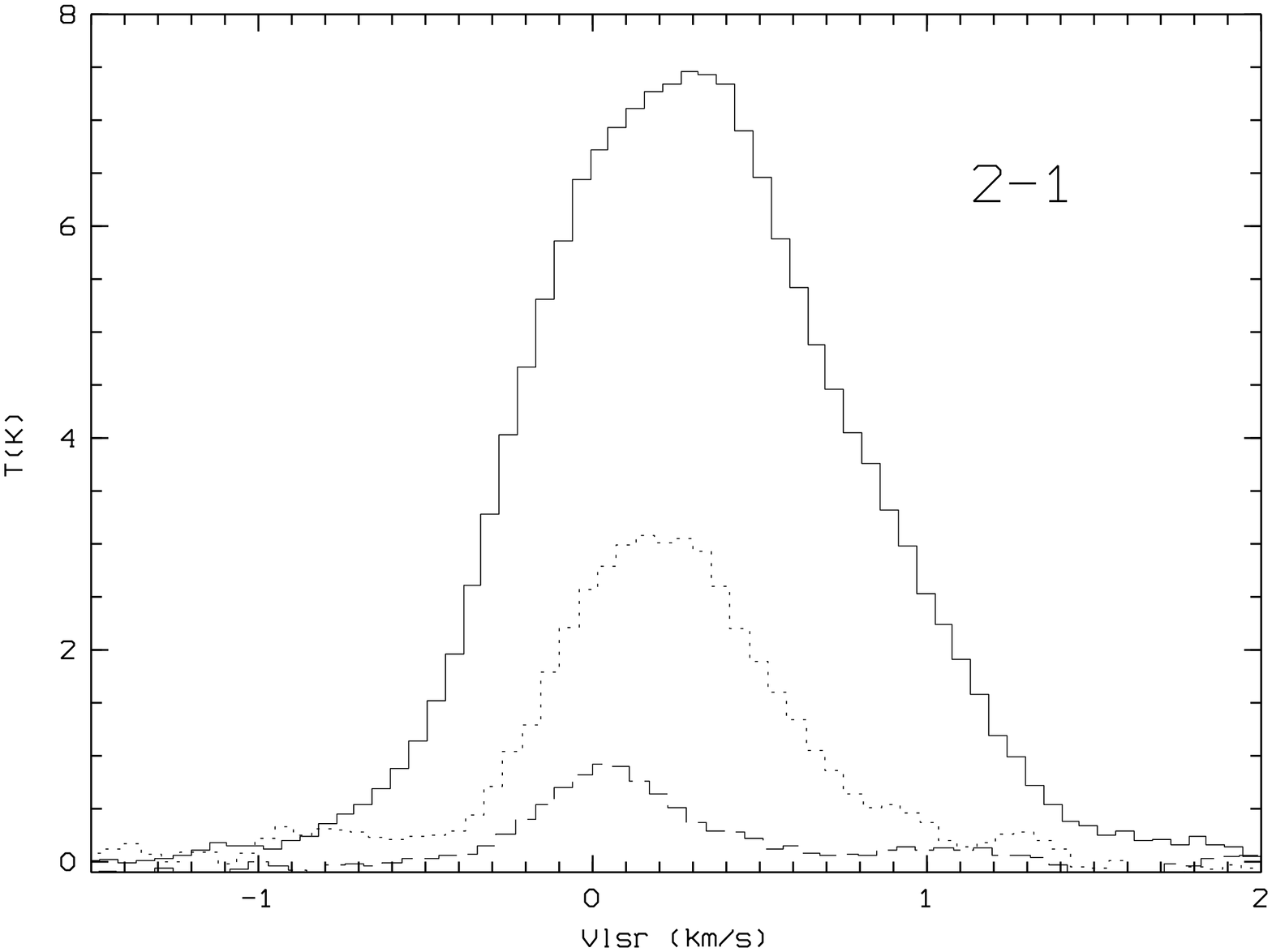,width=8cm,angle=270}
\caption[]{
Averaged spectra observed towards the centre
of L1642 (04h35m10.1s, -14$\degr$15'58'' (J2000)).
The solid, dotted and dashed
histograms are respectively the $^{12}$CO, $^{13}$CO and C$^{18}$O profiles.}
\label{fig:spectra}
\end{figure}

Fig.~\ref{fig:spectra} shows examples of observed line profiles. Typical noise
levels were $\Delta T_{\rm rms}=$0.06\,K for the C$^{18}$O(2--1) spectra and
$\sim$0.15\,K for all the other transitions.

\begin{figure}
\psfig{file=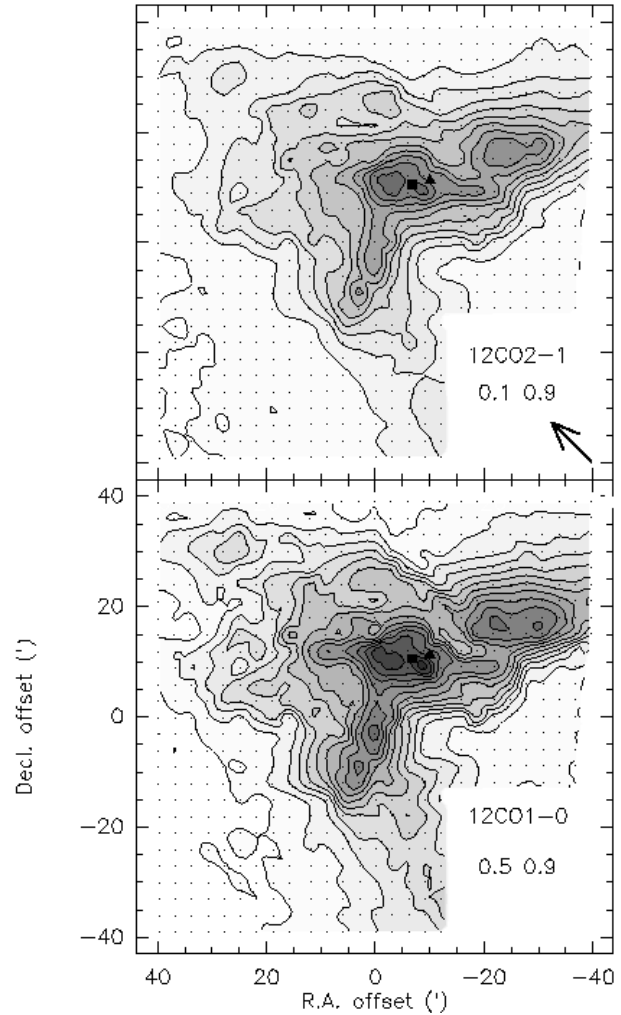,width=8.cm}
\caption{Contour maps of the $^{12}$CO line area. The square and the triangle
indicate
the position of IRAS~04327-1419 and IRAS~04325-1419, respectively. The two numbers
given for each map are the isocontour starting values and the steps
in K\,km\,s$^{-1}$.
The centre position (0,0) is 04h35m30s, -14$\degr$24$\arcmin$28$\arcsec$ (J2000).
The arrow indicates the direction of the cloud tail seen in HI and IRAS maps }
\label{fig:quat12co}
\end{figure}

\begin{figure}
\psfig{file=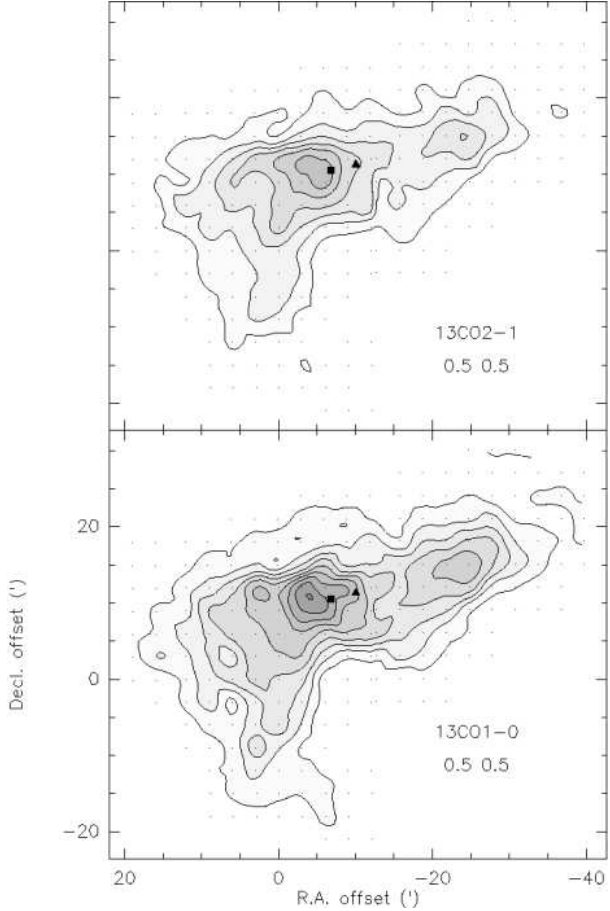,width=8.cm}
\caption[]{As in Fig.\ref{fig:quat12co}, but for $^{13}$CO}
\label{fig:quat13co}
\end{figure}

\begin{figure}
\psfig{file=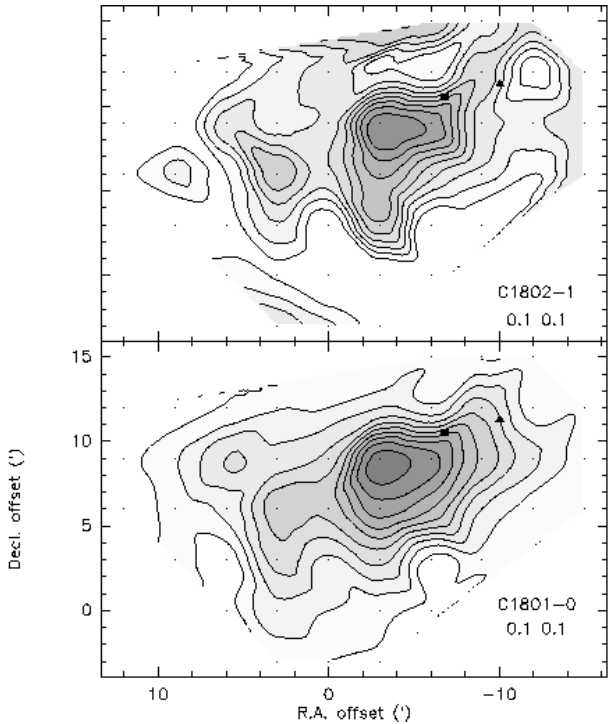,width=8.cm}
\caption[]{As in Fig.\ref{fig:quat12co}, but for C$^{18}$O}
\label{fig:quatc18o}
\end{figure}

\section{Morphological and kinematic analysis} \label{sect:morphology}

The observed line area maps are presented in Figs.~\ref{fig:quat12co}-\ref{fig:quatc18o}.
We utilize the multi-line information to probe the cloud morphology and
kinematics from the dense central parts, as traced by the C$^{18}$O, to the
diffuse outer parts, as traced by $^{12}$CO. The conventional channel maps as
well as the results from principal component analysis (PCA; see e.g. Murtagh
\& Heck \cite{MH87}) are given in Appendix~\ref{app:other}. Here we study
the cloud structure by using Positive Matrix Factorization (PMF), since
comparison with the other methods showed that this method gives the most
compact presentation for the complex cloud structure. The different analysis
methods were not only used to study this particular cloud, Lynds 1642, but
also to evaluate the relative merits of the methods. These are discussed in
Appendix~\ref{sect:comparison}.

\subsection{Positive Matrix Factorization} \label{sect:PMF}

The Positive Matrix Factorization (PMF) method assumes 
the spectra to be the sum of a few basic velocity components. The line
decomposition computes the shape and intensity of each component imposing positivity 
to their profile.
The application of PMF to the analysis of
molecular lines was largely discussed by Juvela et al.(\cite{juvela96}) and we repeat
here only some of the main points. Let $X$ be the data matrix and $\Sigma$ a
matrix consisting of estimated standard deviations of the elements of $X$. Both
matrices have dimensions $n\times m$. Given matrices $X$ and $\Sigma$ and
the rank of factorisation, $r$, PMF calculates a factorisation
\begin{equation}
X = GF + E, G_{ik}\geq 0, F_{kj} \geq 0
\end{equation}
with $i$=1,...,$n$, $k$=1,...,$r$ and $j$=1,...,$m$). $E$ is the residual
matrix and factor matrices $G$ and $F$ are selected to minimise its weighted
norm
\begin{equation}
Q= \sum_{i,j} (E_{i,j}/\Sigma_{i,j})^2 = \sum_{i,j}(\frac{(X-GF)_{ij}}{\Sigma_{ij}})^2.
\end{equation}
The rank $r$ equals the number of spectral components used in approximating
the observed data.

There are two essential improvements over PCA analysis (see
Appendix~\ref{sect:PCA}). Firstly, the elements of the factor matrices, $G$
and $F$ are required to be positive. In this case, this is a physically
meaningful assumption. The observed spectra are decomposed into positive
spectral components and a direct interpretation of the results will be easier
than in the case of PCA, where profiles may contain as many negative as
positive values. Secondly, the residuals are weighted according to individual
error estimates of the observations. This makes it possible to use all
observations without the risk that results will be unduly affected by
individual, low signal-to-noise ratio spectra. Like PCA, the PMF method regards
the observations as a sum of emission components at fixed radial velocities.
In the case of continuous velocity gradients, this is not an exact description
of the data, and this must be kept in mind when interpreting the results.

\begin{figure*}
\resizebox{!}{!}{\includegraphics{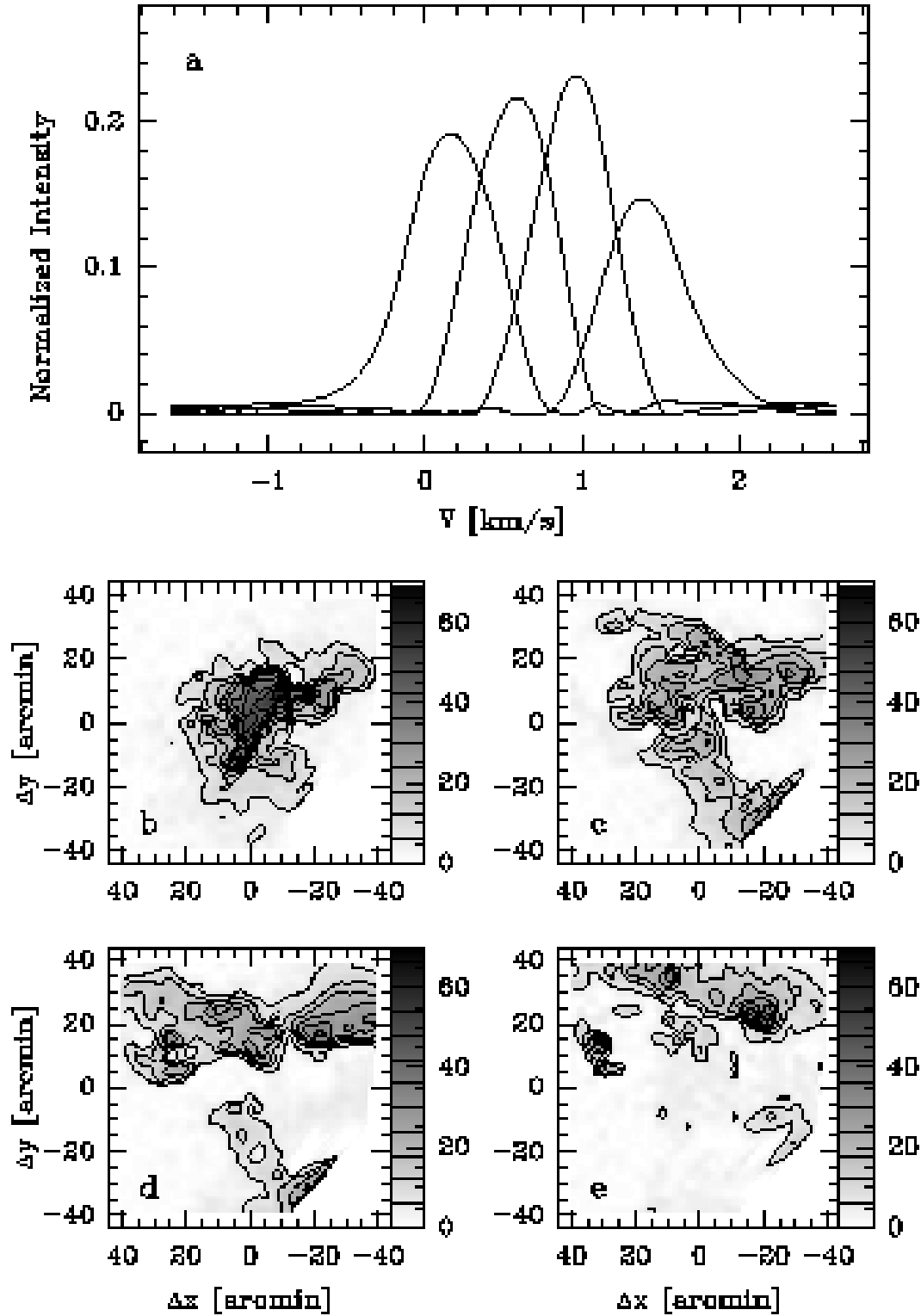}}
\caption[]{Result of the PMF factorisation of $^{12}$CO(1--0) observations
with four spectral components. The spectral profiles of the components are
shown in the first frame. The other frames show the intensity distribution of
the components over the observed map. The component maps are in the order of
increasing radial velocity of the component}
\label{fig:pmf_12co10_4}
\end{figure*}

\begin{figure*}
\resizebox{!}{!}{\includegraphics{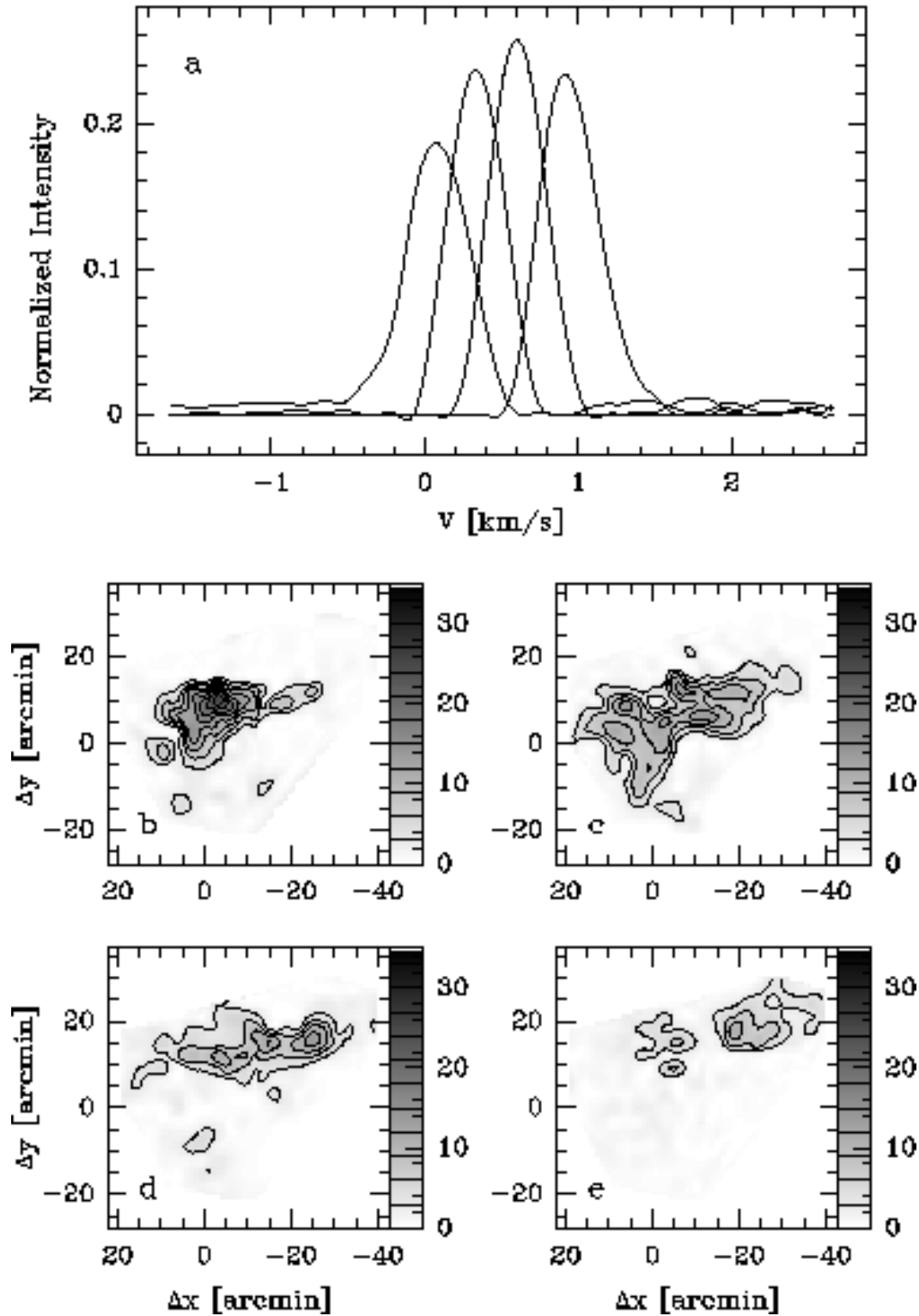}}
\caption[]{Result of the PMF factorisation of $^{13}$CO(1--0) observations
with four components
}
\label{fig:pmf_13co10_4}
\end{figure*}

\begin{figure*}
\resizebox{!}{!}{\includegraphics{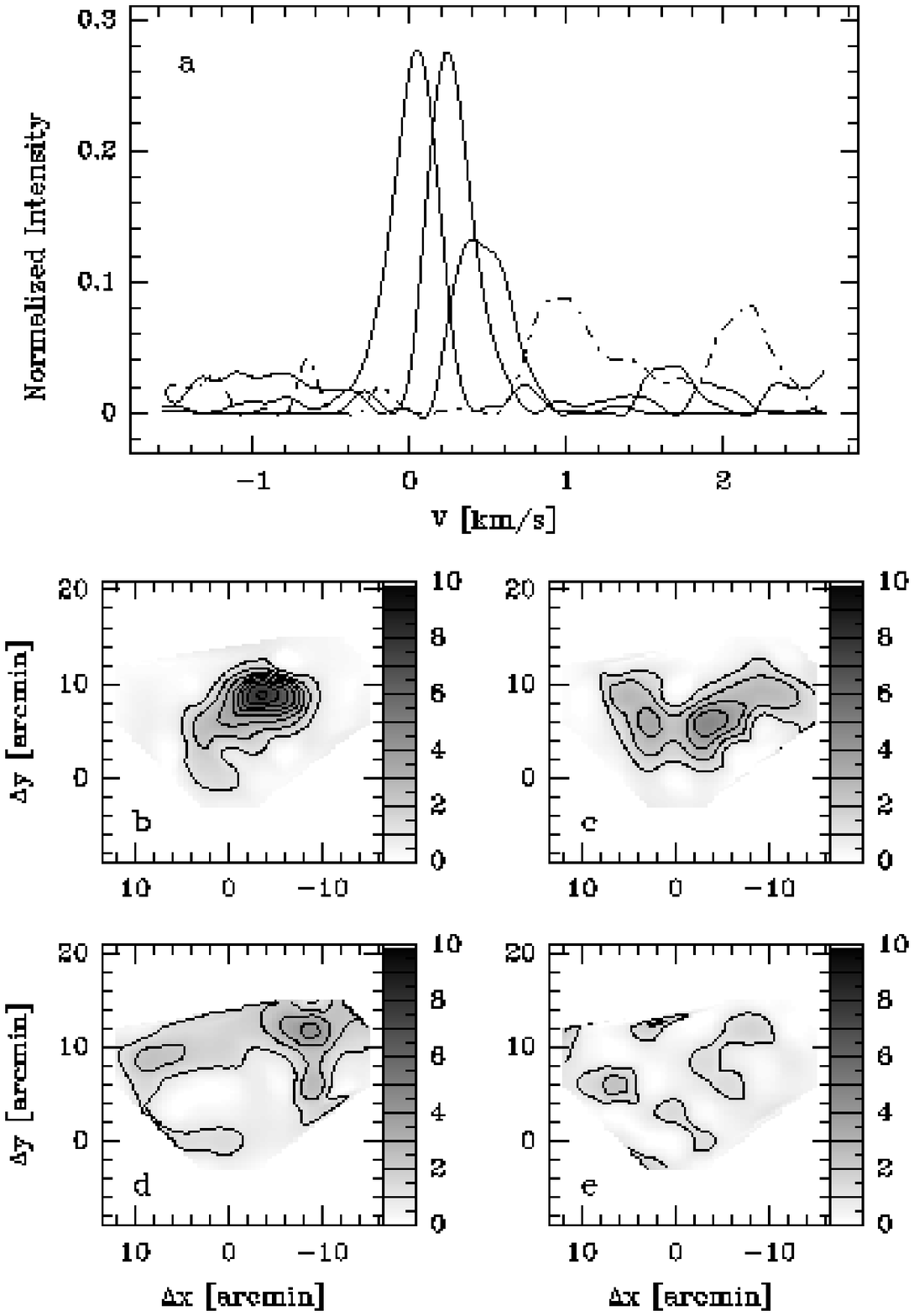}}
\caption[]{Result of the PMF factorisation of C$^{18}$O(1--0) observations
with four components. The last component is mostly noise and is drawn
with dashed line
}
\label{fig:pmf_c18o10_2}
\end{figure*}

PMF requires the selection of the number of components, $r$, and a suitable
value of $r$ could be determined e.g. based on the eigenvalues of matrix $X$.
In practise, it is better to try different $r$-values and select the best
value based on the $Q$-values obtained and the appearance of the components
and the residual maps. The PMF method was run independently on the
$^{13}$CO(1-0), $^{13}$CO(2-1), $^{12}$CO(1-0), $^{12}$CO(2-1), C$^{18}$O(1-0)
and C$^{18}$O(2-1) datasets to extract kinematic information for L1642.
Factorisations were computed for ranks $r$=1 to $r$=5 using the Multilinear
Engine program (Paatero \cite{paatero99}). Results of factorisations with
$r=4$ are shown in Figs.~\ref{fig:pmf_12co10_4}--~\ref{fig:pmf_c18o10_2}. The
figures show the computed spectral profiles of the components for $J$=1-0
lines and their spatial distribution over the mapped area. The results for the
$J$=2-1 transitions were very similar to those of $J$=1-0.
By trying decomposition with 1 to 5 components, we found that, in $^{12}$CO
and $^{13}$CO, 4 components are enough to describe the main kinematic
structures. For C$^{18}$O, only three significant components were found, but
4 components are shown for the sake of completeness.

\subsection{Cloud kinematics and environmental effects} \label{sect:kinematics}
We analyse the kinematic structure of L1642 using the PMF results shown in
Figs.~\ref{fig:pmf_12co10_4}-\ref{fig:pmf_c18o10_2}. The maps of the
kinematic components show an intricate structure. Most of observed CO lines
are not pure gaussians, but exhibit small shoulders and slightly flattened
tops (see Fig.~\ref{fig:spectra}). The possible causes for such line
asymmetries are: ejection, contraction, rotation, self absorption, saturation,
or superposition of several kinematically separate cloud fragments along the
line of sight.

\subsubsection{$^{12}$CO lines}

With $r=4$, the first component of PMF decomposition traces the central part
of L1642. The second one indicates strong emission south of the centre,
extending {\bf to the south west edge of the map}. 
This is still visible in the third
component. However, the last two components mostly trace emission from the
northern part, indicating a velocity gradient towards the north, or possibly
the presence of another cloud component at a higher radial velocity. The
decomposition shows patchy emission at $\sim$1.25\,km\,s$^{-1}$, most notably
the prominent clump at (-20$\arcmin$,+20$\arcmin$) embedded in more extended
emission at $\sim$0.4 km\,s$^{-1}$. With more components ($r$=5), the
$Q$-value improves further by more than 15\%, and decompositions show more
details of the velocity structure in the northern part.
The general kinematic structure is clear, but it cannot be interpreted
directly e.g. as simple rotation, contraction or expansion.
The smallest-velocity component 1, centred at $\sim$0.2\,km\,s$^{-1}$,
delineates the maximum CO emission core of the cloud. At higher radial
velocities (components 2 and 3) emission in both northern and
southern parts of the maps can be seen. The highest-velocity component 4 is
only present
in the extreme North.

Its velocity difference of $\sim$1.3\,km\,s$^{-1}$ with respect to the cloud
core corresponds to a distance of $\sim$1\,pc travelled in 10$^6$ years, a
typical timescale for cloud dynamics.

The residual maps of PMF factorizations with $r$=4 (and with $r$=5) are very
similar for the two $^{12}$CO transitions, and both show several clear clumps,
most notably at positions (0$\arcmin$, 9$\arcmin$) and (-21$\arcmin$,
21$\arcmin$). There is also a wider area of large residuals between these
positions, at around (-10$\arcmin$, +10$\arcmin$), i.e. around the position of
the source IRAS~04325-1419, which is associated with an outflow (Liljestr\"om
et al.~\cite{liljestrom89}).

\begin{figure}
\resizebox{7.8cm}{!}{\includegraphics{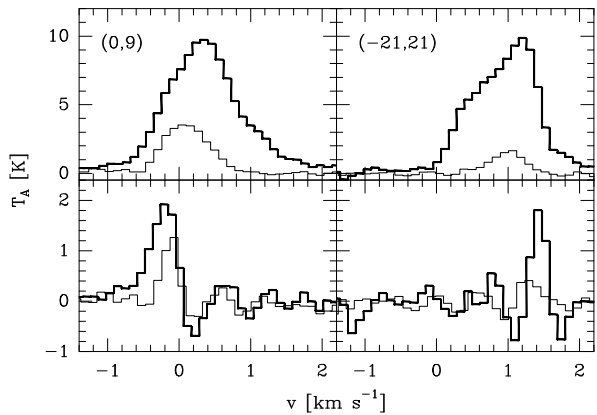}}
\caption[]{Observed $^{12}$CO(1--0) and $^{13}$CO(1--0) spectra (upper frames)
and the residuals (lower frames) of the PMF fit at positions (0$\arcmin$,
9$\arcmin$) and (-21$\arcmin$,21$\arcmin$). The $^{12}$CO spectra are drawn with
thicker lines
}
\label{fig:residuals}
\end{figure}

The study of residuals is a very powerful tool for finding deviating spectra.
By subtracting the decomposition from the observation matrix one can not only
identify deviating spectra, but also see exactly how the profiles differ. For
example, two distinct structures were observed in the PMF residual maps of
both $^{12}$CO(1--0) and $^{12}$CO(2--1). Around position (0$\arcmin$,
9$\arcmin$) the spectra show clear blueshifted excess
($v\sim$-0.5\,km\,s$^{-1}$), while around position (-21$\arcmin$, 21$\arcmin$)
the excess is as clearly redshifted ($v\sim$+1.3\,km\,s$^{-1}$) (see
Fig.\ref{fig:residuals}). These residual features are quite strong, with peak
$^{12}$CO(1--0) antenna temperature above 1\,K. Corresponding features can even be
detected in channel maps. However, the PMF analysis indicates that
the emission seen in the -0.5\,km\,s$^{-1}$ channel is not a normal tail of
the stronger emission at higher velocities. The spectra at the given position
have distinctly different profiles.

\subsubsection{$^{13}$CO lines}

The $^{13}$CO emission is less extended, but it exhibits similar kinematics as
$^{12}$CO.
Figure \ref{fig:pmf_13co10_4} shows the $^{13}$CO(1-0) decomposition with
$r$=4. The maps for the transition $J$=2-1 are again very similar. The first
component is concentrated in the cloud centre, while the last one traces
emission in the north, in particular the two clumps at
(-20$\arcmin$,~+20$\arcmin$) and (-5$\arcmin$,~+15$\arcmin$), already seen in
the $^{12}$CO maps. On the other hand, the ridge of emission south of the
centre is weak in $^{13}$CO. For $^{13}$CO(1-0), a change from $r$=1 to $r$=2
decreases $Q$ by 35$\%$, while the drop is only 15$\%$ in the case of
$^{13}$CO(2-1). A similar difference is noted for higher $r$-values. This is
caused by the difference in the signal-to-noise ratio, and with $r>3$, some
noise ripples appear in the $^{13}$CO(2-1) components. As in the case of
$^{12}$CO residual maps, the $^{13}$CO(1-0) maps show significant residual
emission, e.g. around position(0$\arcmin$,+9$\arcmin$) (see
Fig.~\ref{fig:residuals}).

\subsubsection{C$^{18}$O lines}

C$^{18}$O probes the deepest and densest parts of L1642. Compared with the
previous lines, it is spatially much more concentrated and is restricted to a
velocity range between -0.3 and 0.5\,km\,s$^{-1}$. The PMF 1-component
decomposition of the C$^{18}$O(1-0) shows that the emission has a mean
velocity of 0.2\,km\,s\,$^{-1}$ and a width of 0.47\,km\,s$^{-1}$. We show in
Fig.~\ref{fig:pmf_c18o10_2} the decomposition with
$r=4$ for completeness, even if decompositions with $r>3$ have some irregular
components that are caused mostly by noise. For the two transitions, the maps of
the first three components agree with each other and these are therefore
likely to be real emission features. At low radial velocities, a core located
at (-4$\arcmin$,~+9$\arcmin$) is the main feature. At intermediate velocities,
two separate clumps appear at (+3$\arcmin$,~+5$\arcmin$) and
(-4$\arcmin$,~+6$\arcmin$). The third component shows some emission north of
the core, and, in the maps of both transitions, weak clumps are seen at
(-9$\arcmin$,~+12$\arcmin$) and (+7$\arcmin$,~+9$\arcmin$). The last
components have completely different distributions and spectral profiles and,
and therefore are unlikely to be related to actual cloud structure. Like residual
maps, these `noise factors' can be used to check for systematic errors or
artifacts in data. For example, in the case of C$^{18}$O(2-1), the fourth
spectral component is almost flat and positive, suggesting bad baseline fits
in a few spectra. The two IRAS sources do not coincide with the main peak of
the C$^{18}$O emission, but are located inside the western clump seen in the
third component map, i.e. around velocity $\sim$0.3\,km\,s$^{-1}$.

In addition to the clumps already mentioned, there are separate patches of
emission that are seen in the second and the third component maps. They can be
either independent clumps or parts of a more continuous density distribution
containing gas motions along the line of sight. The latter hypothesis seems
more likely, because the features do not have a clear counterpart on peak
intensity and line area maps.

\subsubsection{Summary of kinematics and environmental effects}

To summarise, PMF (with increasing $r$) provides a hierarchical view of the
cloud kinematics. Firstly, the main cloud is distinguished from the emissions in
the north. Secondly, kinematic substructures appear in the northern part, and
finally, kinematic substructures are identified inside the central area of
L1642. The northern part is likely to belong to L1642, since the velocity
difference is less than the typical velocity dispersion of molecular clouds
(the typical cloud-to-cloud velocity dispersion is 7$\pm$1.5\,km\,s$^{-1}$,
according to Blitz and Fich, \cite{blitz83}).
The $^{12}$CO observations show a systematic velocity structure with the main
emission at $\sim$0.2\,km\,s$^{-1}$ (a velocity close to that of the C$^{18}$O
core), and
fainter and patchy emission with velocities up to 1.6\,km\,s$^{-1}$ extending
to the north and to the south.

\begin{figure}
\resizebox{!}{!}{\includegraphics{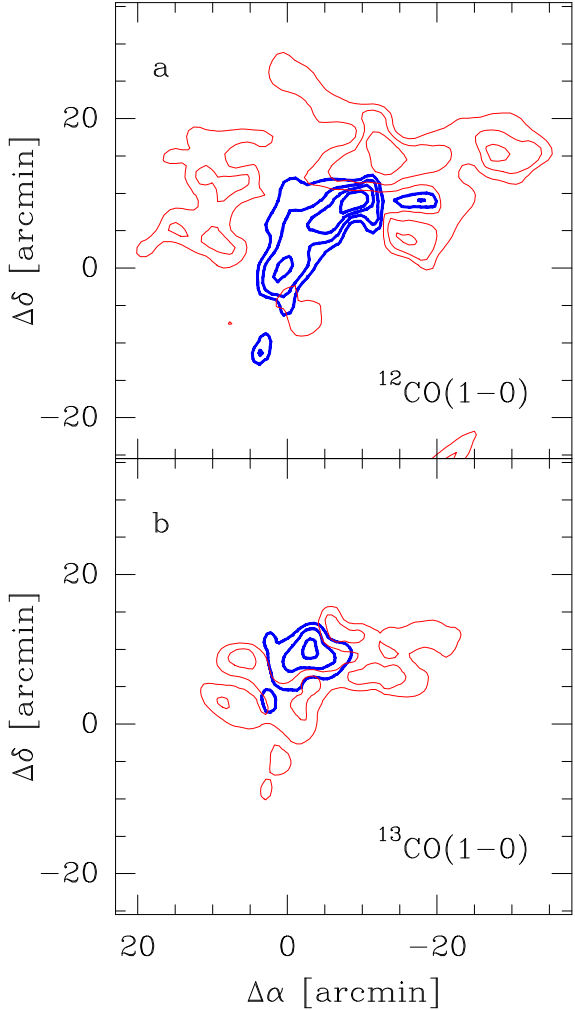}}
\caption[]{The spatial distributions of the first two PMF components
in the case of the $^{12}$CO(1-0) and the $^{13}$CO(1-0) observations. The
blue and red contours correspond to the first and the second PMF component,
respectively, as shown in Figs.~\ref{fig:pmf_12co10_4} and \ref{fig:pmf_13co10_4}. The
contour levels are 40, 50, and 55 for the first and 20, 25, and 30 for the
second $^{12}$CO component, and 14, 19, and 24 for the first and 8.5 and 11.5
for the second $^{13}$CO component. The units are the same as used in
Figs.~\ref{fig:pmf_12co10_4} and \ref{fig:pmf_13co10_4}. }
\label{fig:redblue}
\end{figure}

In the PMF decompositions of the $^{12}$CO lines the first component shows the
compact core. The second component shows almost a ring-like structure at
$\sim$0.5\,km\,s$^{-1}$ higher radial velocity. This is illustrated in
Fig.~\ref{fig:redblue}a where we have overlayed the contours of the first two
PMF components. Similar structure is seen in the decomposition of the
$^{13}$CO(1-0) observations (see Fig.~\ref{fig:redblue}b). This suggests some kind
of an expansion. At higher velocities, no emission is seen towards the centre
of the cloud. Therefore, the structure cannot be spherically symmetric, but it
could be an incomplete expanding shell, the far side of which is
missing.

The L1642 cloud is the head of a large cometary structure with a tail
extending more than 5 degrees in the north eastern direction. The tail is seen
very clearly, e.g. in the 100$\mu$m IRAS map. Furthermore, HI data shows a
velocity gradient along the tail, with increasing velocity towards the
north-east (Taylor et al. \cite{taylor82}). This suggests that the first PMF
component represents the very head of the whole structure, and gas in the head
is flowing relative to the rest of the structure, not only towards the
south-west but also more towards us. The second PMF component would correspond
to more diffuse material being left behind by the moving cloud core. In the
north-eastern corner, the spatial separation of the third and fourth PMF
components indicates a velocity gradient across the tail. 
This velocity gradient closely agrees with the HI velocity structure
(Liljestr\"om \& Mattila \cite{liljestrom88}; Taylor et al. \cite{taylor82})
indicating that the atomic and molecular gas components are well mixed in the
cloud.

In projection on the sky, L1642 is located in the direction of the edge of the
Orion-Eridanus Bubble, and we must consider the possibility of their interaction.
>From multi-wavelength analysis, Reynolds and Odgen
(\cite{reynolds79}) have shown that the Orion-Eridanus Bubble is an expanding
cavity (expansion velocity $\sim$15\,km\,s$^{-1}$) full of warm ionised gas
surrounded by an expanding HI shell. X-ray enhancements have been measured and
attributed to hot gas flowing out from the bubble through a chimney or a hole
where L1642 would be located (Heiles et al. \cite{heiles99}). From HI data
(Brown et al. \cite{brown95}), the bubble velocities range between -40 and
+40\,km\,s$^{-1}$. The largest spatial extension occurs between -1 and
+8\,km\,s$^{-1}$, which is the most probable range for the systematic velocity
of the bubble. L1642 falls in the velocity range of the Orion-Eridanus Bubble,
which suggests that it could have been swept up and hence shaped by the
bubble. This would favour the hypothesis that the tail is formed as a
stationary cloud embedded in moving interstellar medium. On HI maps (Brown et
al. \cite{brown95}), L1642 appears isolated for velocities between -1.5 to
2.5\,km\,s$^{-1}$. At these velocities, the bubble edges delineate an `empty
space' roughly centered on L1642. More precisely, L1642 is kinematically
disconnected from the filaments at $l\sim$210$\degr$ which have velocities
less than -3\,km\,s$^{-1}$.

\section{Correlations of the line intensities} \label{sect:correlations}

\begin{figure}
\psfig{file=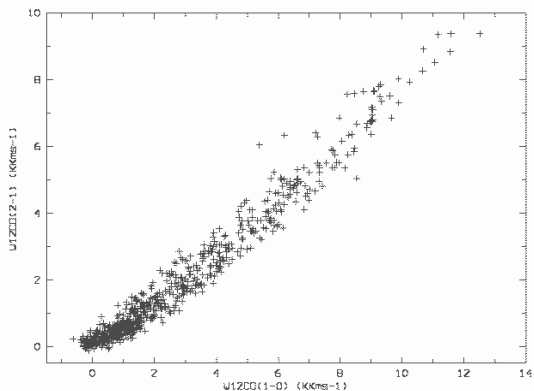,width=8cm,angle=270}
\caption[]{Correlation between $^{12}$CO $J=$2--1 and $J=$1--0 line areas}
\label{fig:12co_profile}
\end{figure}

We use the multi-transition information to examine the excitation conditions
of L1642. There is a good correlation between the integrated line intensities
of the $J=$2--1 and $J=$1--0 transitions (Fig.~\ref{fig:12co_profile}). The
following relations are obtained for the line areas:
\begin{eqnarray}
W[^{12}{\rm CO(2-1)}] &= (0.75\pm0.06) \,\,W[^{12}{\rm CO(1-0)}] \nonumber \\ &- (0.12\pm0.40)
\nonumber \\
W[^{13}{\rm CO(2-1)}] &= (0.71\pm0.01) \,\,W[^{13}{\rm CO(1-0)}] \nonumber \\ &- (0.01\pm0.19)
\nonumber \\
\end{eqnarray}
These relations are in agreement with other high latitude clouds (e.g. Ingalls
et al. \cite{ingalls00}). The $^{12}$CO/$^{13}$CO ratio is much smaller than the
typical relative abundance of the species, which is a natural consequence of
$^{12}$CO being optically thick.

\begin{figure}
\psfig{file=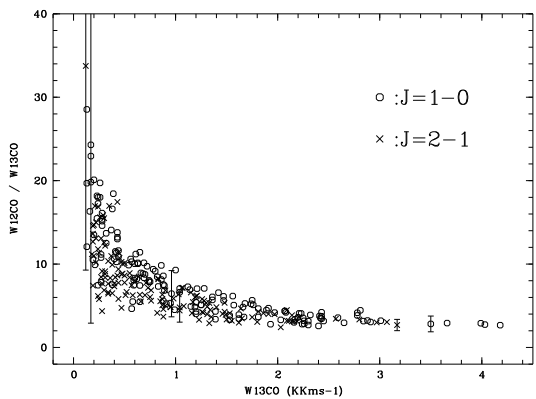,width=8.cm,angle=270}
\caption[]{ W[$^{12}$CO]/W[$^{13}$CO] versus W[$^{13}$CO]}
\label{fig:area_ratio}
\end{figure}

The decrease of $W[^{12}$CO]/W[$^{13}$CO] versus W[$^{13}$CO]
(Fig.~\ref{fig:area_ratio}) is caused by the saturation of the CO line. The
few high ratio points at low $W[^{13}$CO] are attributed to the case in which
both isotopes are optically thin and the asymptotic value of about 3 is
reached when $^{13}$CO becomes optically thick. These small ratios are found
in the central parts of the cloud.

The mean value of the ratio $W$[$^{13}$CO(1-0)]/$W$[C$^{18}$O(1-0)] is
7.2$\pm$0.3. Within a central region of some 10$\arcmin$$\times$15$\arcmin$
the ratio decreases, presumably due to $^{13}$CO saturation, below the
normally assumed isotopic abundance ratio of 5.5. On the other hand,
everywhere in the outer region of the area covered by C$^{18}$O observations
the ratio increases to almost 10. Although $^{13}$CO emission is still not
completely optically thin, the observed variation also indicates a change in
the relative abundance of the two species. The $^{13}$CO/C$^{18}$O ratio is
enhanced to values $>$5.5 by $^{13}$CO fractionation. The average ratio for
the corresponding $J=$2--1 lines is 6.8$\pm$2.2.

\section{Column density distribution and mass of L1642} \label{sect:colden}

\subsection{LTE-analysis}

In order to calculate the column density, one must first
determine the excitation temperature throughout the cloud. The standard
methods assume LTE conditions, and either $^{12}$CO is assumed to be optically
thick, or the optical depth ratio of two isotopic lines is assumed equal to the
terrestrial abundance ratio (the assumed $^{12}$CO to $^{13}$CO abundance ratio is
$\sim$89). In addition, an identical excitation temperature is usually assumed
for both isotopes.

In our case, both methods agree up to distances of $\sim$30$\arcmin$ from the
cloud centre (the adopted cloud center is: -5$\arcmin$, +10$\arcmin$), after
which the $^{13}$CO antenna temperature values are affected by noise. 

\begin{figure}
\psfig{file=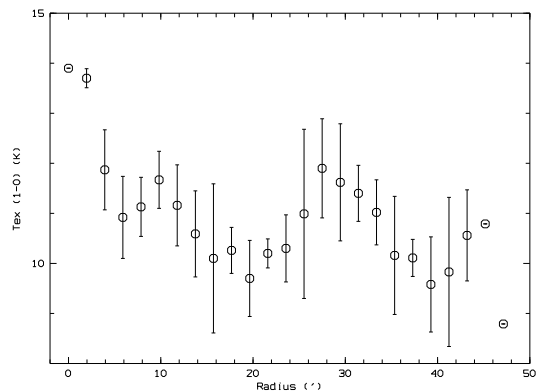,width=8.cm,angle=270}
\caption[]{Excitation temperature $T_{\rm ex}$(1--0), calculated from 
$^{13}$CO/$^{12}$CO peak antenna temperature ratio, versus radius. 
Error bars reflect dispersion in each distance bin}
\label{fig:radial_tex}
\end{figure}

The excitation temperature has a nearly constant value from cloud centre
to edge (Fig.~\ref{fig:radial_tex}).

In the central part of the cloud, $^{12}$CO and $^{13}$CO emission lines are
certainly optically thick, and hence the derived temperature may
not be representative of the bulk of the cloud. The best temperature estimate
for the cloud centre is obtained from the $^{13}$CO/C$^{18}$O ratio which
gives a value of $T_{\rm ex}$=8\,K. Based on the optically thick $^{12}$CO
emission, the mean temperature estimate is $T_{\rm ex}$=10.7\,K.

The column densities and the cloud mass were estimated based on the $^{13}$CO
data. The estimates are derived from the $J$=1--0 line, since according to
radiative transfer models (see Appendix~\ref{appendix:lte}) this gives more
reliable estimates than the $J$=2--1 transition.

\begin{figure}
\psfig{file=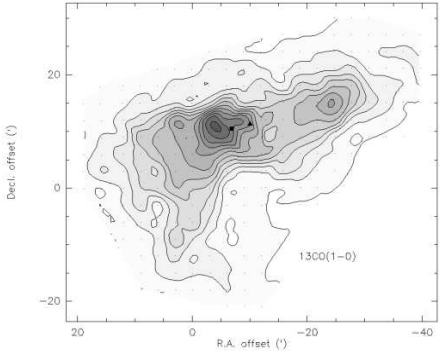,width=8cm,angle=270}
\caption[]{Map of H$_{2}$ column density derived from $^{13}$CO(1-0) lines
assuming LTE. The contours range from 1.0 10$^{20}$ to 5.6 10$^{21}$ cm$^{-2}$
by 5.0 10$^{20}$ cm$^{-2}$. The filled square and the triangle indicate the
positions of IRAS~04237-1419 and IRAS~04325-1419}
\label{fig:n2den}
\end{figure}

\begin{figure}
\psfig{file=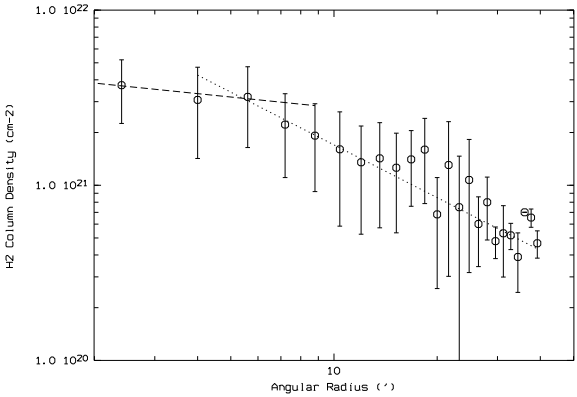,width=8cm,angle=270}
\caption[]{The profile of the H$_2$ column density derived from $^{13}$CO(1-0)
observations. The dotted lines are for r$^{-1}$ trends and dashed lines for
r$^{-0.2}$. The centre is at $\Delta \alpha$=-5$\arcmin$, $\Delta
\delta$=+10$\arcmin$ (see Fig.~\ref{fig:n2den})} \label{fig:n2prof}
\end{figure}

For LTE calculations, we assume a constant excitation temperature of 10K.
Figure~\ref{fig:n2den} shows the H$_{2}$ column density map derived from
$^{13}$CO(1--0). The adopted conversion factor is N(H$_{2}$)/
N($^{13}$CO)=1\,10$^{6}$.  IRAS~04237-1419 is close to the column density
maximum, while IRAS~04325-1419 is located half way to the edge of the central
core. The H$_2$ column density was averaged in concentric rings of angular
radius $r$ from the centre, and the resulting
radial column density distribution is shown in Fig.~\ref{fig:n2prof}. The
vertical error bars represent the dispersion of the values averaged at each
radius. An $r^{-1}$ distribution is obtained in the outer parts of the
$^{13}$CO cloud ($r>$5.6$\arcmin$), but the distribution in the core
($r<$5.6$\arcmin$) is significantly flatter. Such behaviour is commonly
observed in cloud cores (Bacmann et al.,\cite{bacmann00}; Ward-Thompson et
al., \cite{ward94}, \cite{ward99}), and is in good agreement with both the
previous $^{13}$CO observations of L1642 (Liljestr\"om,
\cite{liljestrom91}, see her Fig.~7) and the density profile of dust
(Laureijs et al. \cite{laureijs87}).

The cloud mass can be calculated according to
\begin{equation}
M(M_{\odot })=5.74\,10^{-20}\,d_{100pc}^{2}\int \int N(H_{2})\cos \delta
\,d\delta \,d\alpha ,
\end{equation}
where $d_{100pc}$=$d$/100\,pc, with $d$ the distance of the cloud. Within the
boundary $T_{\rm R}^{\ast }(^{13}{\rm CO}(1-0))\geq 1$\,K, corresponding to
$H({\rm H}_2) \ge 6 \times 10^{20}$\,cm$^{-2}$, we obtain from the $N$(H$_2$)
map as shown in Fig.~\ref{fig:n2den} $M=(30\pm 10)\,M_{\odot
}\,d_{100pc}^{2}\sim 59\,M_{\odot }$ (assuming $d$=140 pc).

\subsection{Spherically symmetric non-LTE models}

In reality, the excitation temperature is not likely to be constant over the
whole line of sight. This may lead to erroneous results in the normal LTE
analysis, especially at low temperatures (e.g. Padoan et al.
\cite{padoan00}). For this reason, we compared the LTE results, as described
above, with non-LTE radiative transfer calculations. The comparison gives an
estimate for the uncertainty of the results. Precise determination of 
the cloud mass requires detailed knowledge of the density and temperature
distributions.

The $^{13}$CO observations were modelled with a spherically symmetric cloud
with radial density distribution $n\sim r^{-2.0}$, except for the central 10\%
of the cloud radius for which constant density was adopted. The density
contrast between the centre and the cloud surface was set to 20. The radiative
transfer problem was solved with a Monte Carlo method (Juvela
\cite{juvela97}), and the results were compared with observed spectra which
were averaged over concentric annuli. The density and the size of the model
cloud were varied in order to find parameter values that best reproduced the
observations. A value of 1.0$\cdot$10$^{-6}$ was assumed for the fractional
abundance of $^{13}$CO.

Isothermal models with $T_{\rm kin}$ between 8 and 14\,K were first
considered. The $\chi^2$ values were practically independent of the
temperature and it is clear that the temperature cannot be well constrained
based on $^{13}$CO observations alone. The mass estimates ranged from
82\,M$_{\sun}$ at 8\,K to $\sim$65\,M$_{\sun}$ at 14\,K. The intensity ratio
between the 2-1 and 1-0 lines was generally too low and, as expected for a
microturbulent model, the line profiles toward the centre positions were
strongly self-absorbed. 

\begin{figure}
\resizebox{\hsize}{!}{\includegraphics{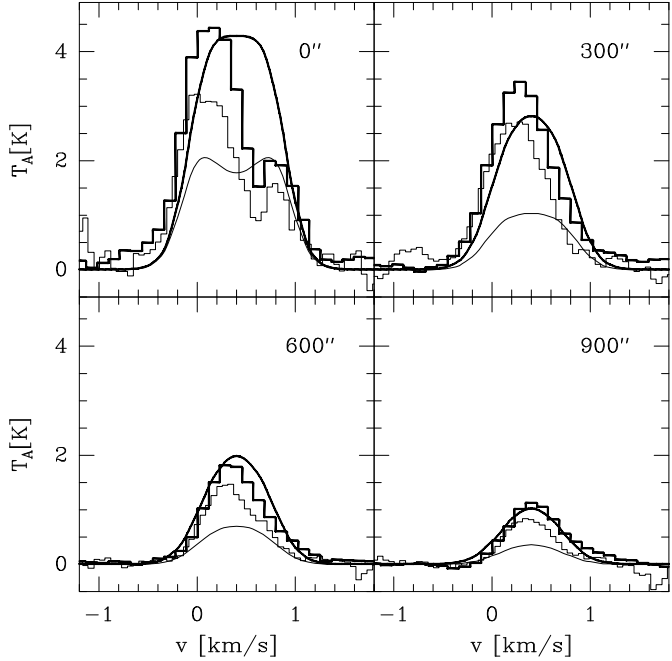}} 
\caption[]{
Spectra from a spherically symmetric model cloud (solid lines) and observed
spectra (histograms) averaged over different annuli centred at the position
4h35m13.6s -14$\degr$15$\arcmin$28$\arcsec$. In the model the kinetic
temperature increased linearly from 9\,K in the centre to 14\,K on the cloud
surface. The radius of the annulus is given in each frame. Thick lines
represent $^{13}$CO(1-0) and thin lines $^{13}$CO(2-1) spectra.}
\label{fig:1dfit}
\end{figure}

We further studied a set of models in which the kinetic temperature increase
linearly with the radius. These did not lead to significantly better fits but
the mass estimates were slightly lower and close to the LTE values. For a
model where $T_{\rm kin}$ increases from 9\,K in the centre to 14\,K on the
cloud surface, the estimated mass was 61\,$M_{\sun}$ and the central density
was 1.2$\cdot$10$^{4}$\,cm$^{3}$. For this model the fitted spectra are shown
in Fig.~\ref{fig:1dfit}.

As can be seen the spherically symmetric models do not provide a good fit to
the observations. At the centre position the observed spectra are clearly
asymmetric and when observations are averaged over concentric annuli the
presence of different velocity components or gradients leads to artificial
line broadening or even multiple peaks in the averaged profiles. Such effects
can be taken into account only in three-dimensional models (see
Sect.~\ref{sect:3d} add Appendix~\ref{appendix}). The main shortcoming of the
spherically symmetric models is, however, the low line ratio between the
$J$=2-1 and $J$=1-0 transitions. This could indicate either a much higher
kinetic temperature, significant foreground absorption or more likely density
inhomogeneities inside the source.

The observed average intensity ratio $^{13}$CO(2-1)/$^{13}$CO(1-0) is 0.69
(the average of observations with W($^{13}$CO(1-0))$>$1\,K\,km\,s$^{-1}$).
Assuming LTE conditions, constant excitation along the line of sight, and
an excitation temperature of $T_{\rm ex}$=10\,K (Fig.~\ref{fig:radial_tex}),
such a ratio would imply an optical depth of $^{13}$CO(2-1)$\approx$2. 
The $^{13}$CO molecule is, however, not thermalized. In our calculations, the
excitation temperature of the 2--1 transition is always clearly below that of
the transition 1--0.  For example, in the centre of the model cloud with
$T_{\rm kin}$=12K the difference is almost three degrees. At higher densities
$^{13}$CO would become less subthermally excited and the ratio
$^{13}$CO(2-1)/$^{13}$CO(1-0) would increase. Therefore, the observed high
line ratios may indicate that the cloud is inhomogeneous and $^{13}$CO is
partly thermalized in spite of the low average density.

\subsection{Three-dimensional non-LTE models} \label{sect:3d}

The line profiles from a real, turbulent cloud are seldom as clearly
self-absorbed as in the microturbulent models. The cloud was therefore also
modelled with fully three-dimensional models. In these isothermal models,
the general density distribution was based on the LTE column density maps.
Details of the calculations are given in Appendix~\ref{appendix:3d}. The best
fit was obtained at a significantly lower temperature than for the spherical
homogeneous models, $T_{\rm kin}$=8\,K, leading to a mass estimate of
70\,$M_{\sun}$. However, the temperature cannot be accurately determined, and
for the model with $T_{\rm kin}$=14\,K, the average line ratio,
$^{13}$CO(2-1)/$^{13}$CO(1-0)=0.63, is still reasonably close to the observed
value of 0.69. For any given temperature, the predicted mass values were only
slightly lower than in the spherical models.

We applied a standard LTE analysis to the spectra produced by the 3D models.
This analysis showed that, in the parameter range considered here, the LTE
column density estimates based on the 1--0 are reliable (see
Appendix~\ref{appendix:lte}). The mass estimate from the LTE-calculations are
indeed consistent with the 3-D modelling, when a kinetic temperature of
$T_{\rm kin}\sim 12\,$K is assumed.

\subsection{Dynamical state of the cloud core}

We investigate the dynamical state of L1642, following the method by
Liljestr\"om (\cite{liljestrom91}), who estimated the ratio $\frac{\sigma_{\rm
gas}}{\sigma_{\rm vir}}$ where $\sigma_{\rm gas}$ and $\sigma_{\rm vir}$ are
the velocity dispersion of the gas derived from observation and the virialized
velocity dispersion of the mean gas particle, respectively. The effective
radius of the cloud is $R_{\rm eff}=0.97\,d_{100pc}$\,pc ($R_{\rm
eff}=\sqrt{ab}$; $a$=semi-major axis, $b$=semi-minor axis).  A value of
$\sigma_{\rm gas}$=0.17\,km\,s$^{-1}$ is obtained from the $^{13}$CO line
width (corrected for instrumental and line opacity broadening). Adopting
$T_{\rm kin}$=12\,K and assuming $d$=140pc, we find $\sigma_{\rm vir,
1-dim}$=0.21$\pm$0.02 and hence a ratio of $\frac{\sigma_{\rm
gas}}{\sigma_{\rm vir}}$=0.81$\pm$0.08 for a homogeneous sphere. For a
centrally condensed sphere with $n \propto r^{-1}$, we obtain ${\sigma_{\rm
vir}}$=0.19$\pm$0.02 and $\frac{\sigma_{\rm gas}}{\sigma_{\rm
vir}}$=0.89$\pm$0.08. The values are very close to 1, suggesting that the
cloud is probably in virial equilibrium. Wed note, that $^{13}$CO
traces only some 80\% of the total cloud mass (e.g. Falgarone and Puget,
\cite{falgarone88}), while another 20\% is in more diffuse cloud medium.

\section{Conclusions} \label{sect:conclusions}

We have presented $^{12}$CO, $^{13}$CO and C$^{18}$O $J$=1--0 and $J$=2--1 line
observations of the translucent high latitude cloud L1642. The main results of
this study are summarized as follows:

- Besides the conventional channel maps, we have also applied the Principal
Component Analysis (PCA) and Positive Matrix Factorization (PMF). The PMF
method was found to produce an easily understandable, compressed presentation
of the data that preserved all the important morphological and kinematic
features of the cloud.

- The morphological and kinematic analysis has shown a main structure at
a velocity of 0.2\,km\,s$^{-1}$ with an incomplete ring structure around it at larger
velocities. Fainter emission extends to the north
with velocities up to 1.6 km\,s$^{-1}$.

This structure suggests an incomplete expanding shell, the front side of which
is seen in the main component.

An elongated morphology in the line of sight direction is also suggested by
the kinematics.

- In C$^{18}$O, only two spatially well separated clumps can be distinguished.
These structures do not coincide with the two embedded IRAS sources, which are
offset by 5--7$\arcmin$ from the main peak.

- We confirm that L1642 follows an $r^{-1}$ density distribution in its outer parts and a
flatter distribution with $n \propto r^{-0.2}$ for the core. The cloud is probably in
virial equilibrium.

- The LTE estimate of the cloud mass is 59\,$M_{\sun}$.

- We performed three-dimensional radiative transfer modelling of the $^{13}$CO
observations and obtained a cloud mass estimate of $\sim$70\,$M_{\sun}$. The
cloud temperature is, however, not well constrained, leading to large
uncertainties in the mass estimates.

\begin{acknowledgements}
We thank the anonymous referee for useful comments on an earlier version of
this paper.

We acknowledge the support from the Academy of Finland Grants no. 1011055,
173727, 174854
\end{acknowledgements}

\appendix

\section{Alternative methods for the study of cloud kinematics} \label{app:other}

\subsection{Channel maps} \label{sect:channel}

\begin{figure*}
\psfig{file=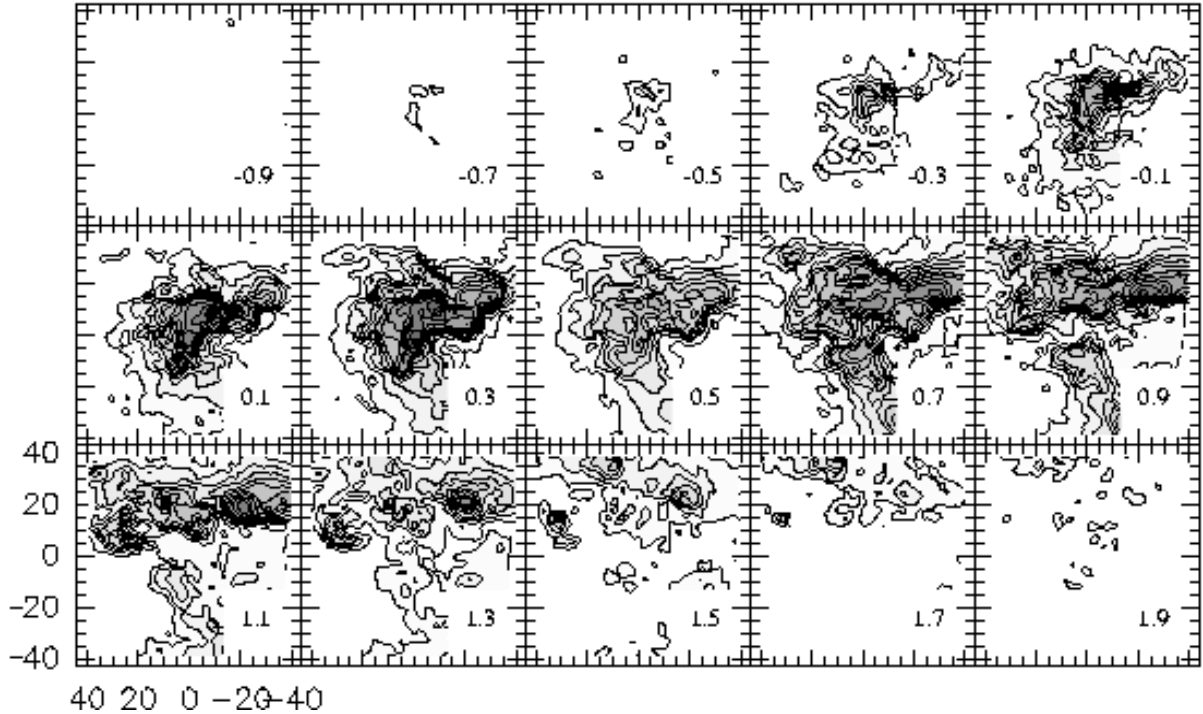,width=16.cm}
\caption[]{$^{12}$CO(1--0) channel maps. The velocity (in \,km\,s$^{-1}$)
is indicated on each panel. The contour interval is 0.3\,K\,km\,s$^{-1}$ and the
first contour is at 0.2\,K\,km\,s$^{-1}$}
\label{fig:12co10chn}
\end{figure*}

\begin{figure*}
\psfig{file=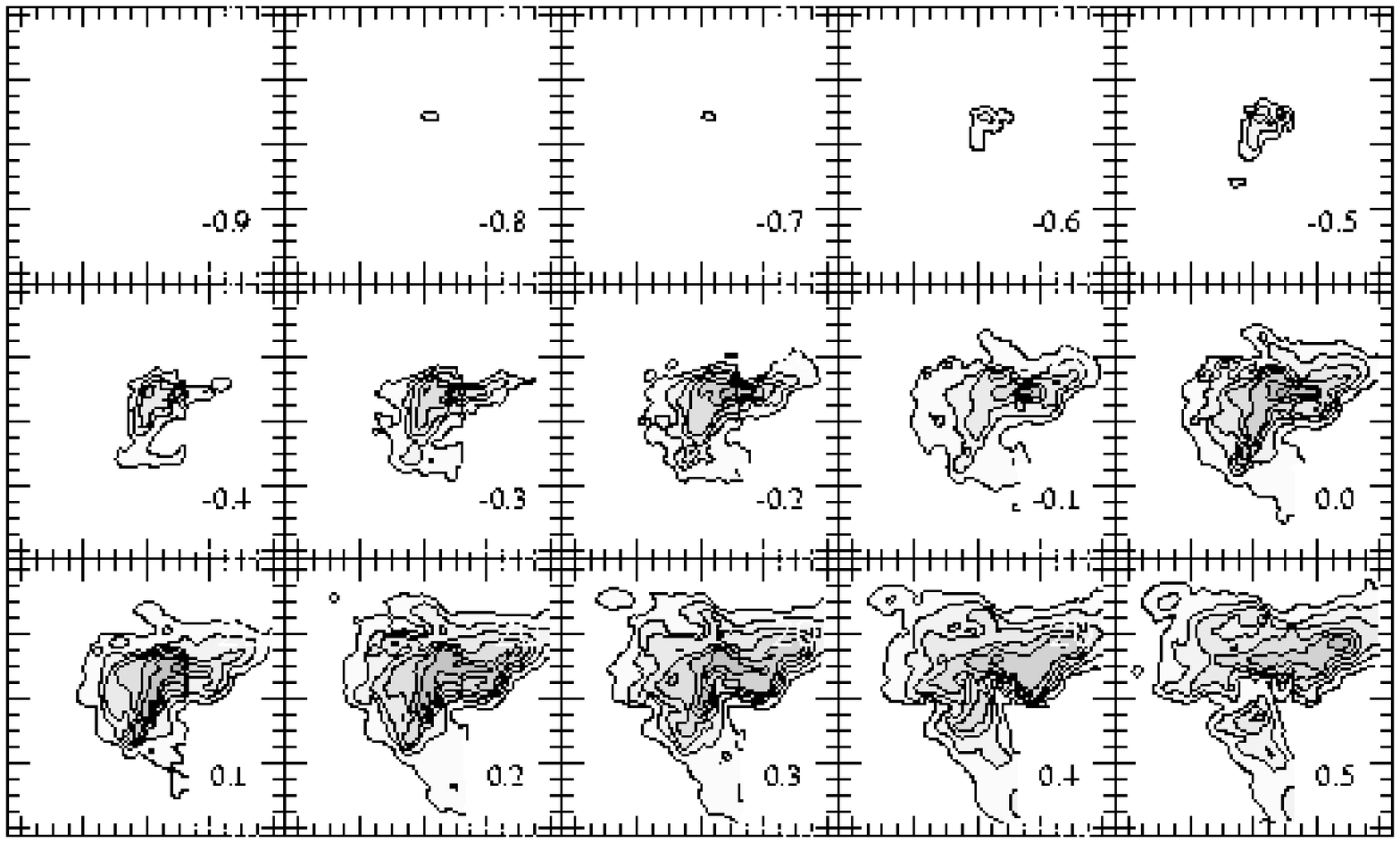,width=16.0cm}
\vspace{-0.71cm}
\psfig{file=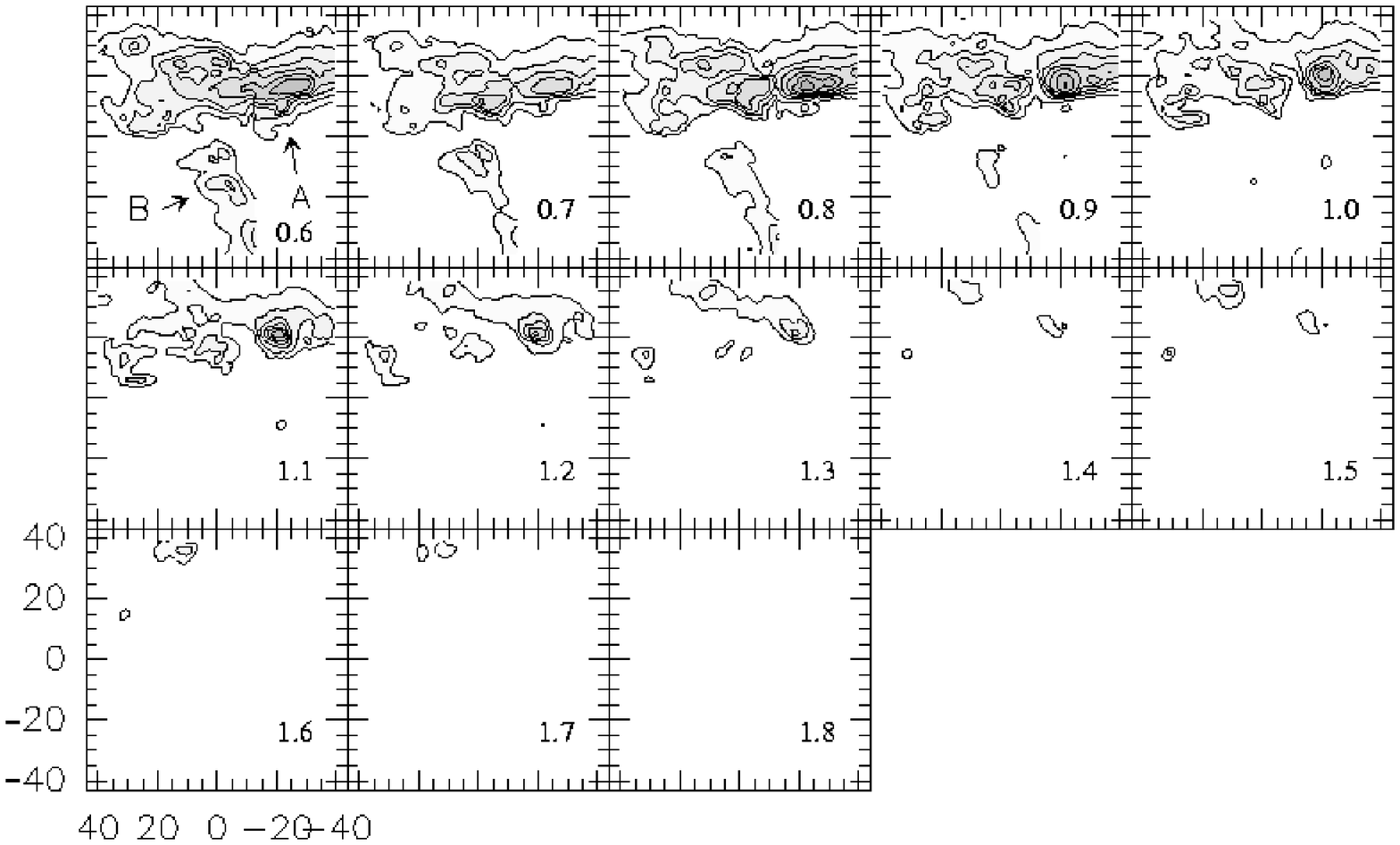,width=16.0cm}
\caption[]{$^{12}$CO(2--1) channel maps.
The velocity (in \,km\,s$^{-1}$) is indicated on each panel. The
contour interval is 0.2\,K\,km\,s$^{-1}$ and first contour is at 0.2\,K\,km\,s$^{-1}$.
On this figure, structures A and B, mentioned in the text,
are indicated on the 0.6\,km\,s$^{-1}$ channel}
\label{fig:12co21chn}
\end{figure*}

\begin{figure*}
\psfig{file=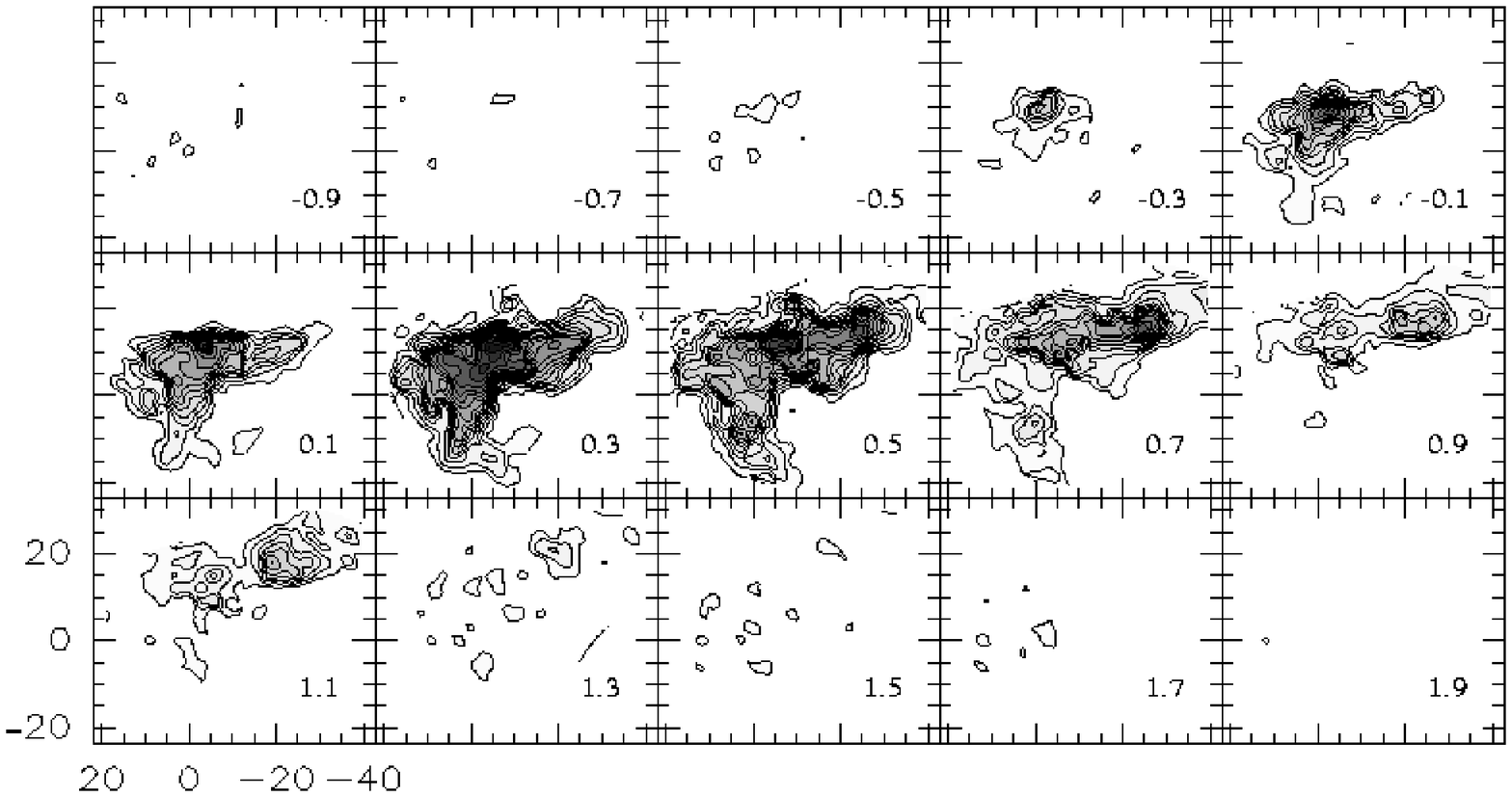,width=16cm}
\psfig{file=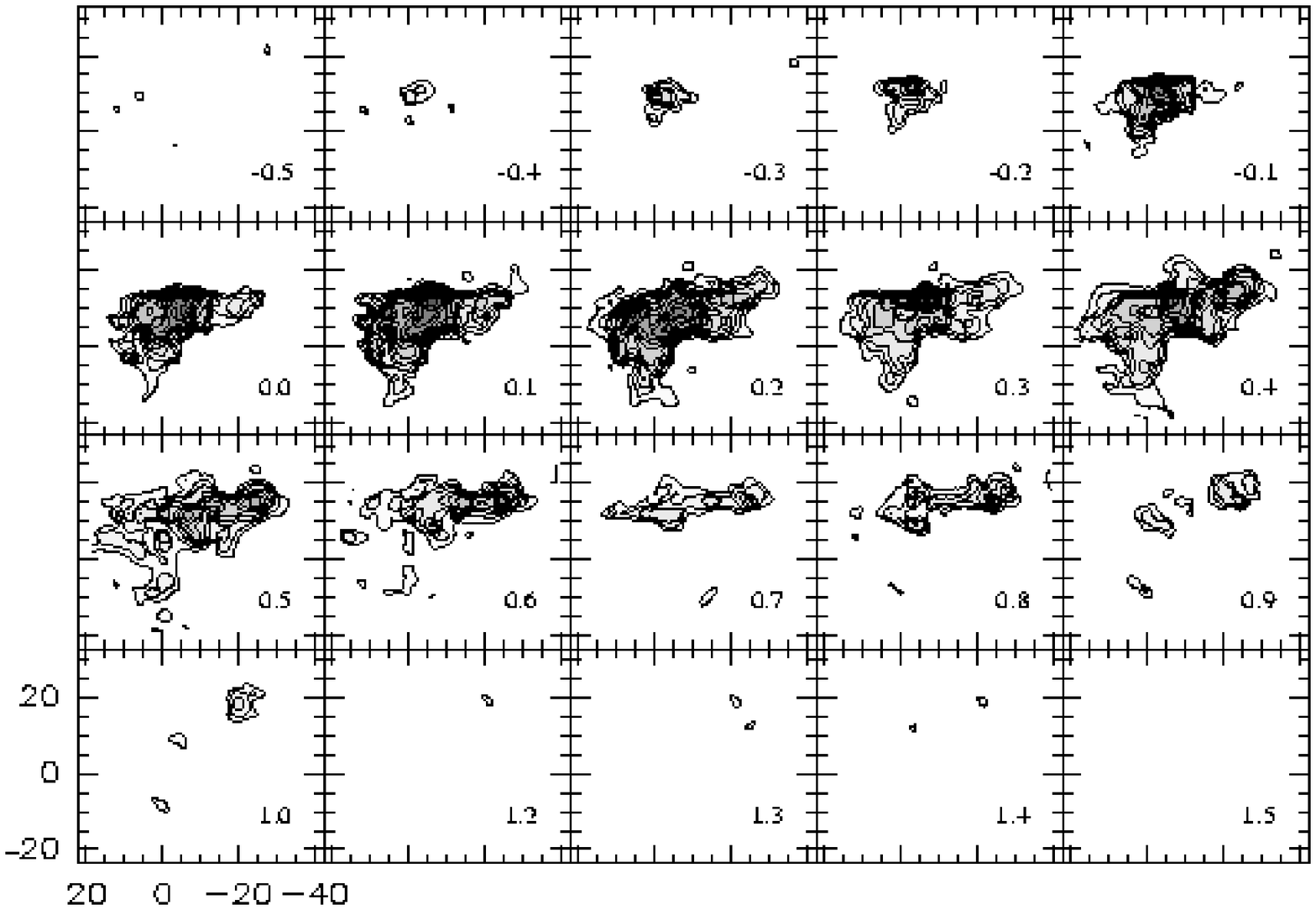,width=16cm}
\caption[]{$^{13}$CO(1--0) (upper) and $^{13}$CO(2--1) (lower) channel
maps. The velocity (in \,km\,s$^{-1}$) is indicated on each panel. The
contour interval is 0.1\,K\,km\,s$^{-1}$ for 1--0 and 0.05 K\,km\,s$^{-1}$ for 2--1, both starting
at 0.1\,K\,km\,s$^{-1}$ }
\label{fig:13co10chn}
\end{figure*}

\begin{figure*}
\psfig{file=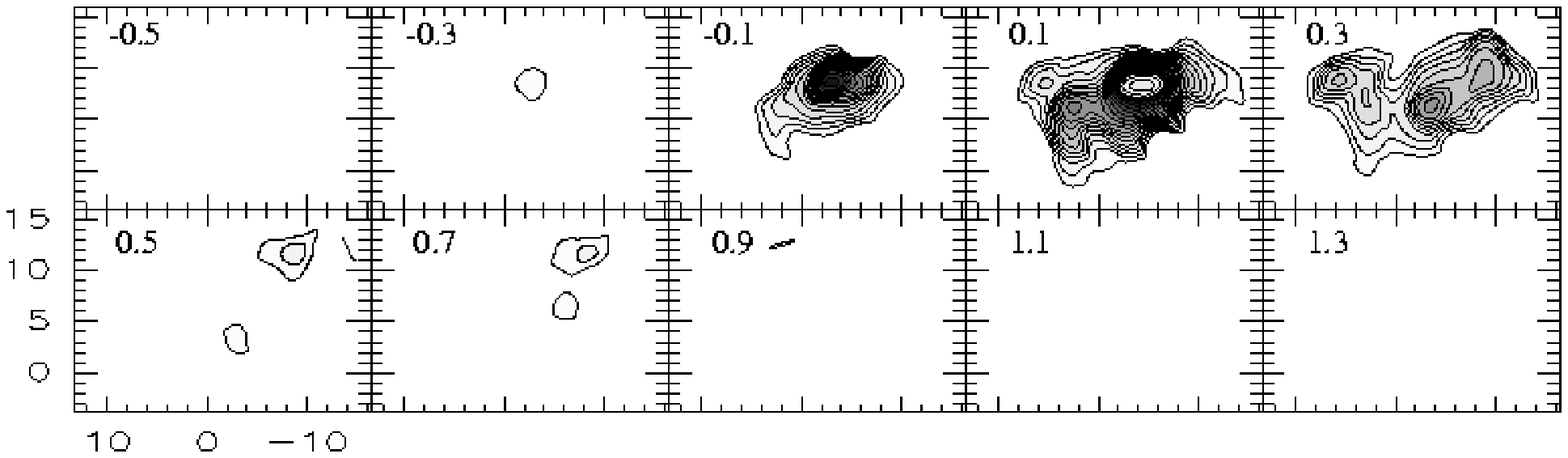,width=16.cm}
\psfig{file=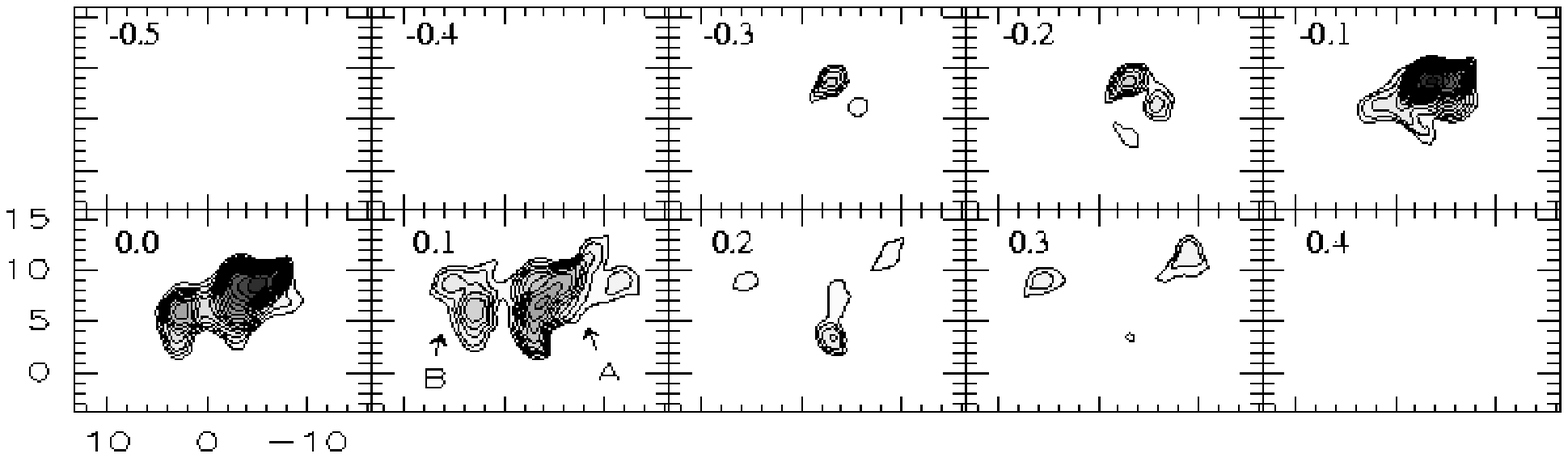,width=16.cm}
\caption[]{C$^{18}$O(1--0) (upper) and C$^{18}$O(2--1) (lower) channel
maps. The velocity (in \,km\,s$^{-1}$) is indicated on each panel.
The contour interval is 0.03\,K\,km\,s$^{-1}$ for 1--0 and
0.02\,K\,km\,s$^{-1}$ for 2--1, the lowest contour lying at 0.1\,K\,km\,s$^{-1}$}
\label{fig:c18o10chn}
\end{figure*}

The $^{12}$CO, $^{13}$CO and C$^{18}$O channel maps are shown in Figs.
\ref{fig:12co10chn} to \ref{fig:c18o10chn}.
In $^{12}$CO, the first significant emission appears at {-0.5}\,km\,s$^{-1}$
around position (-3$\arcmin$, +10$\arcmin$) and, as the velocity increases,
emission extends towards both south and west. Above -0.1 km\,s$^{-1}$, the
general morphology remains the same with only small additional patches
appearing. These suggest the presence of small internal motions and clumpy
structure. Finally, above 0.5\,km\,s$^{-1}$, the cloud breaks up into two
separate structures.  The first one (structure A; see
Fig.~\ref{fig:12co21chn}) is elongated in the east-west direction and is
located around $\Delta y\sim$20$\arcmin$. The second one (structure B) extends
from the centre towards the south.

Structure A remains smooth between 0.6 and 0.9\,km\,s$^{-1}$.  At higher
velocity, it fragments with persistent clumps at (-20$\arcmin$, +20$\arcmin$)
and (+32$\arcmin$, +15$\arcmin$). The latter is well identified in
$^{12}$CO(1--0) but is less clear in $^{12}$CO(2--1). The velocity of these
clumps is $\sim$1.2\,km\,s$^{-1}$.  Structure B fades after 1\,km\,s$^{-1}$,
the mean velocity being $\sim$0.8\,km\,s$^{-1}$.  At higher velocities, the
main structure is a faint elongated emission emanating from the (-20$\arcmin$,
+20$\arcmin$) clump.

\begin{figure}
\psfig{file=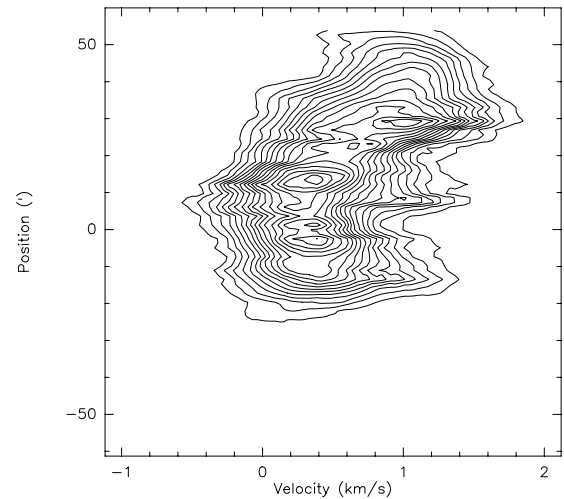,width=8.0cm,angle=270}
\caption[]{Velocity diagram of $^{12}$CO(1--0) observations along
the line from position (40$\arcmin$,-40$\arcmin$) to (-40$\arcmin$,40$\arcmin$).}
\label{fig:vdiag12co10}
\end{figure}

Fig.~\ref{fig:vdiag12co10} shows the velocity diagram for $^{12}$CO(1--0)
observations along the line between positions (40$\arcmin$,-40$\arcmin$) and
(-40$\arcmin$,40$\arcmin$). The average velocity increases from
$\sim$0.3\,km\,s$^{-1}$ in the south to $\sim$1\,km\,s$^{-1}$ at the northern
end where the velocity increases all the way to the edge of the map. The three
preeminent clumps at $\sim$(0$\arcmin$, 0$\arcmin$), (-10$\arcmin$,
+10$\arcmin$) and (-20$\arcmin$, +20$\arcmin$) are visible as separate maxima.
However, only the last one (at offset 30$\arcmin$ in
Fig.~\ref{fig:vdiag12co10}) is clearly separated from the main cloud based on
its radial velocity. The gradual change between offsets 15$\arcmin$ and 30$\arcmin$
does not necessarily require a continuous velocity gradient but could be
caused by the superposition of emission at two separate velocities. The range
of velocities is large, both at the location of the northern clump at
(-20$\arcmin$, +20$\arcmin$) (offset +30$\arcmin$) and in the south (offset
-15$\arcmin$). However, the largest dispersion is found at offset $7\arcmin$
where the emission extends to high positive radial velocities, especially
compared with the velocity of the main emission. The location is some
5$\arcmin$ from IRAS~04325-1419, the source of a known outflow (Liljestr\"om
et al.~\cite{liljestrom89}).

The $^{13}$CO emission is less extended, but it exhibits similar kinematics as
$^{12}$CO. In $^{13}$CO(2--1), map emission clumps are clearly seen around
(+6$\arcmin$, +12$\arcmin$), (-6$\arcmin$, +11.5$\arcmin$) and (+9$\arcmin$,
-2$\arcmin$). The (-20$\arcmin$, +20$\arcmin$) emission clump is still
dominant, with a mean velocity at around 0.8\,km\,s$^{-1}$.

C$^{18}$O probes the deepest and densest parts of L1642. Compared with the
$^{12}$CO and $^{13}$CO lines, it is spatially much more concentrated and is
restricted to a velocity range between -0.3 and 0.5\,km\,s$^{-1}$. The
emission mostly occurs around $\sim$0.15\,km\,s$^{-1}$.  Two main parts can be
distinguished around $\Delta x\sim$-5$\arcmin$ (part A) and $\Delta
x\sim$3$\arcmin$ (part B) (see Fig.~\ref{fig:c18o10chn}). Both have several
patches appearing at different velocities: in A at (-4$\arcmin$, +9$\arcmin$)
($V_{\rm lsr}\sim$-0.1\,km\,s$^{-1}$), (-6$\arcmin$, +9$\arcmin$) ($V_{\rm
lsr}\sim$0.1\,km\,s$^{-1}$), (-4$\arcmin$, +6$\arcmin$) ($V_{\rm
lsr}\sim$0.15\,km\,s$^{-1}$) and (-4$\arcmin$, +3.5$\arcmin$) ($V_{\rm
lsr}\sim$0.25\,km\,s$^{-1}$), and in part B at (+2$\arcmin$, +6$\arcmin$)
($V_{\rm lsr}\sim$0.1\,km\,s$^{-1}$) and (+6$\arcmin$,+9$\arcmin$) ($V_{\rm
lsr}\sim$0.3\,km\,s$^{-1}$). These could either be independent clumps or
velocity cuts of a continuous density distribution where there are gas motions
along the line of sight. The latter hypothesis is more likely, because the
features do not have a clear counterpart on peak intensity and line area maps.

\subsection{Principal component analysis} \label{sect:PCA}

The spectral data consist of an ensemble of $n$ objects (profiles) each with
$p$ attributes (intensity at each velocity channel) and can thus be formally
represented as an $n\times p$ matrix $X$.  The projection of the data onto an
arbitrary axis $u_{\alpha}$ is given by $y_{\alpha}$ = $Xu_{\alpha}$.  The
principal component analysis (PCA; Murtagh \& Heck \cite{MH87}) determines a
set of orthogonal axes $u_{\alpha}$, such that the projection of $X$ upon each
consecutive $u_{\alpha}$ always maximises the variance along that axis.
Mathematically, the base is determined by computing the eigenvectors
$u_{\alpha}$ of the covariance matrix. The method gives an objective way of
determining the characteristic radial velocities of structures which may not
even be discernible upon visual inspection.

The relative importance, $C_{\alpha}$, of the axis $\alpha$ is defined as
$C_{\alpha}=\frac{\lambda_{\alpha}}{\sum_{\alpha}\lambda_{\alpha}}$, i.e. the
eigenvalue $\lambda_{\alpha}$ indicates the percentage of the variance of the
observed variables explained by the corresponding principal component. Usually
two or three factorial axes with the highest $C_{\alpha}$ value are enough to
describe the data. Further details about the application of PCA to the study
of the interstellar medium can be found in Heyer and Schloerb
(\cite{heyer97}), \'Abraham et al. (\cite{abraham00}) and Ungerechts et al.
(\cite{ungerechts97}).

PCA was applied independently to $^{12}$CO(1--0), $^{12}$CO(2--1),
$^{13}$CO(1--0), $^{13}$CO(2--1), C$^{18}$O(1--0) and C$^{18}$O(2--1) data
cubes. For $^{13}$CO and C$^{18}$O lines, the eigenvectors are very irregular.
Normal PCA analysis does not take into account error estimates, and a good
signal-to-noise ratio is important, especially as PCA is computed using
normalised profiles. Therefore, we limited the PCA analysis to the
$^{12}$CO(1-0) data.

\begin{figure}
\psfig{file=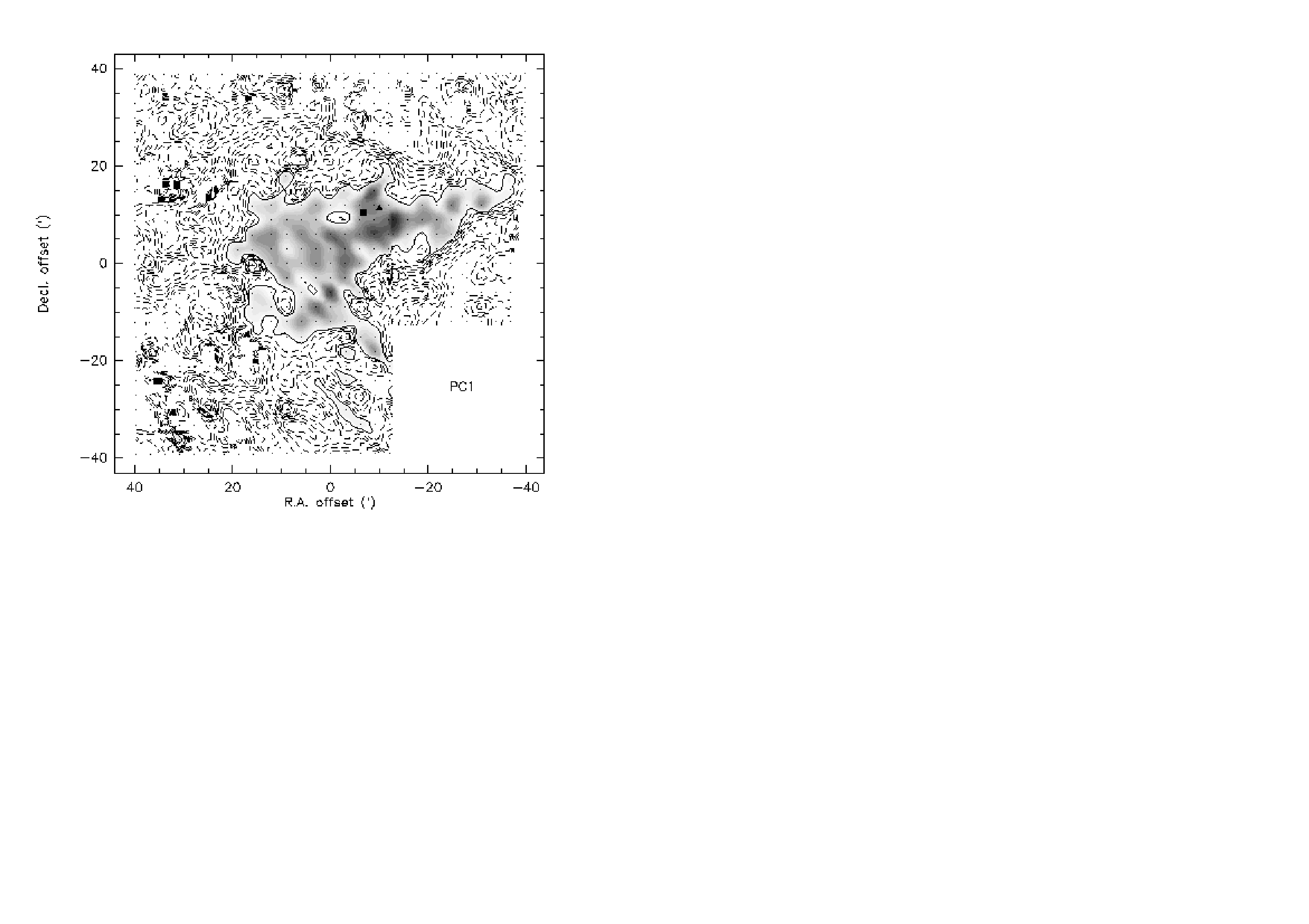 ,width=9cm,height=5.8cm,angle=270}
\psfig{file=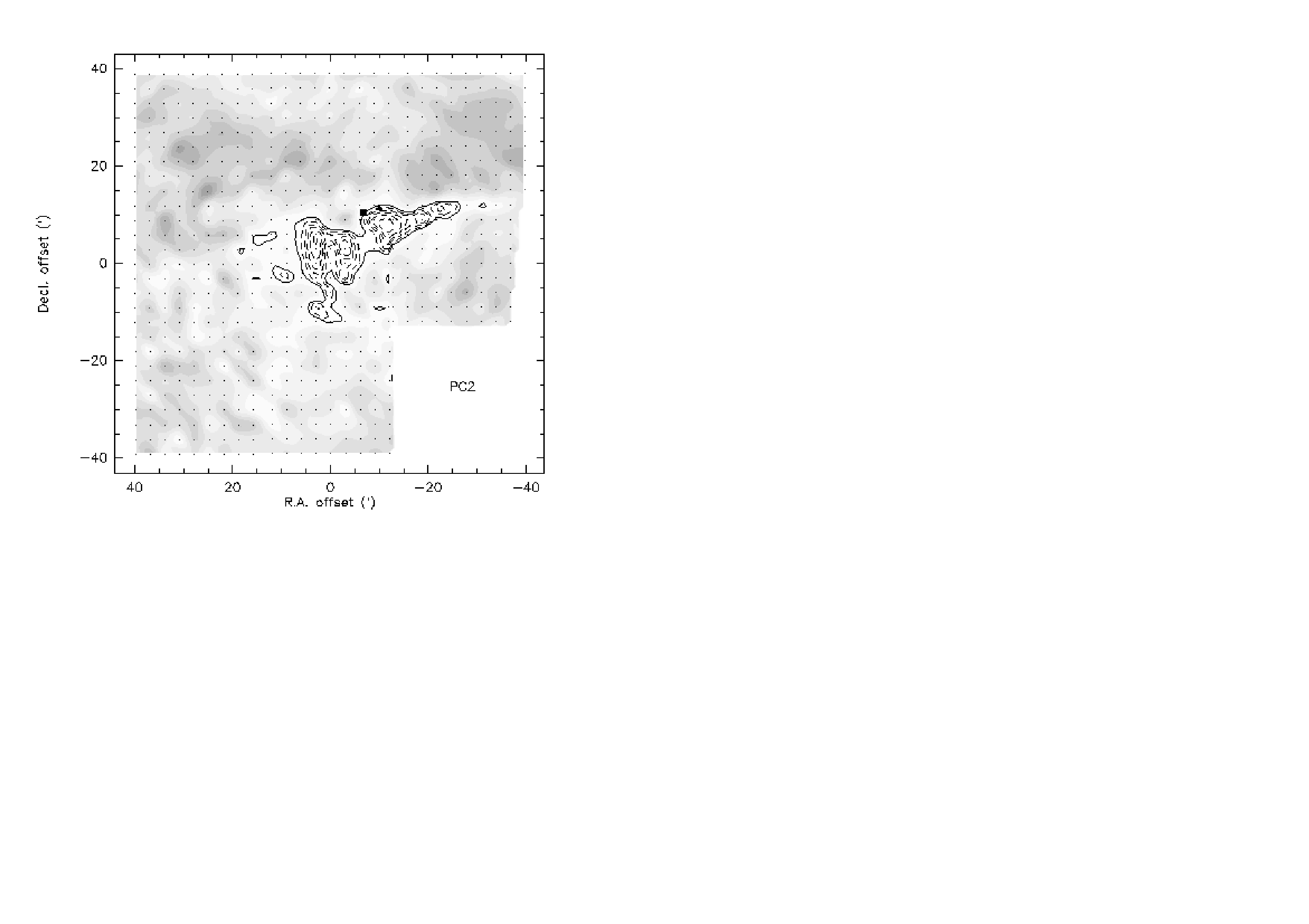,width=9cm,height=5.8cm,angle=270}
\psfig{file=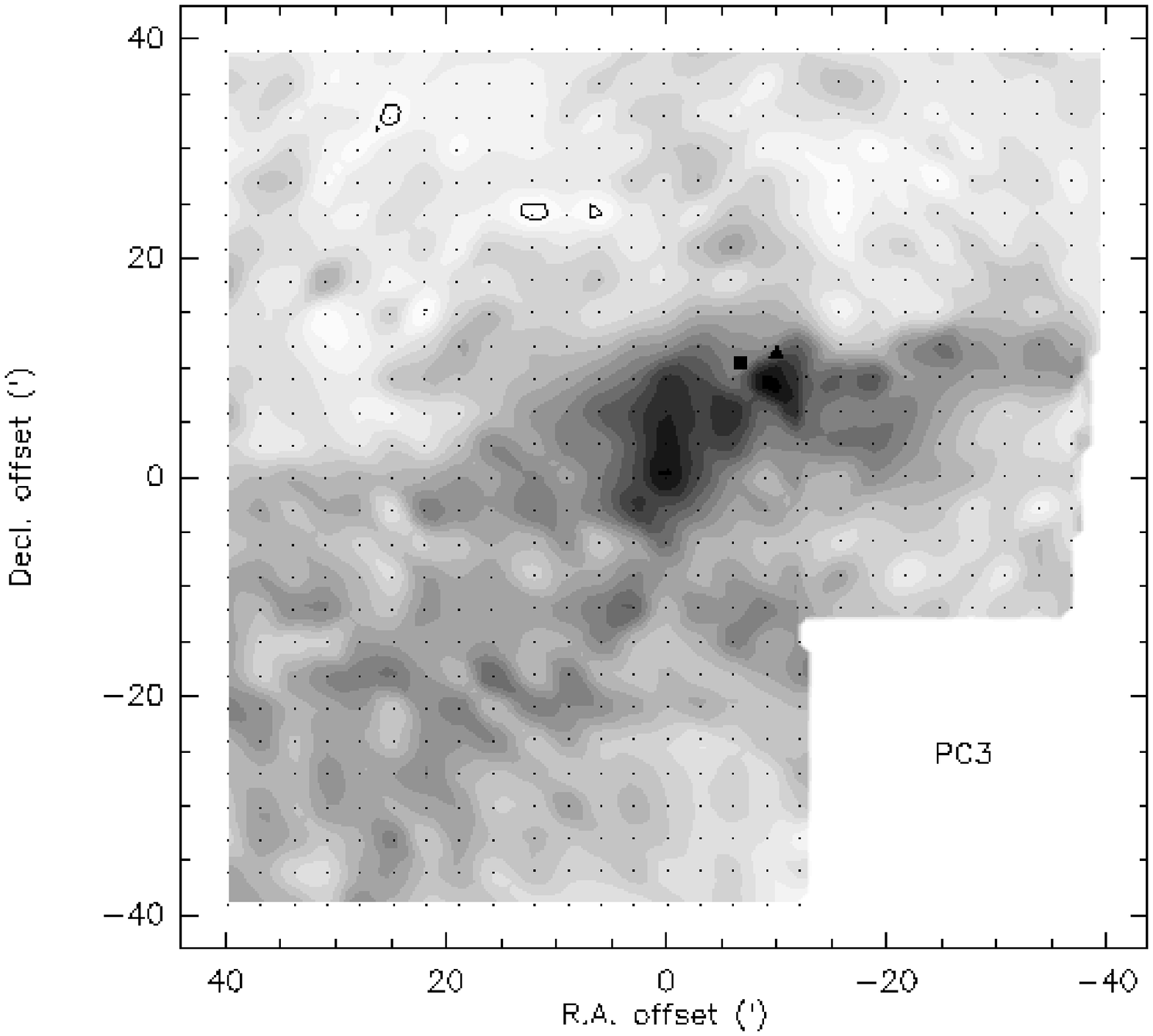,width=9cm,height=5.8cm,angle=270}
\psfig{file=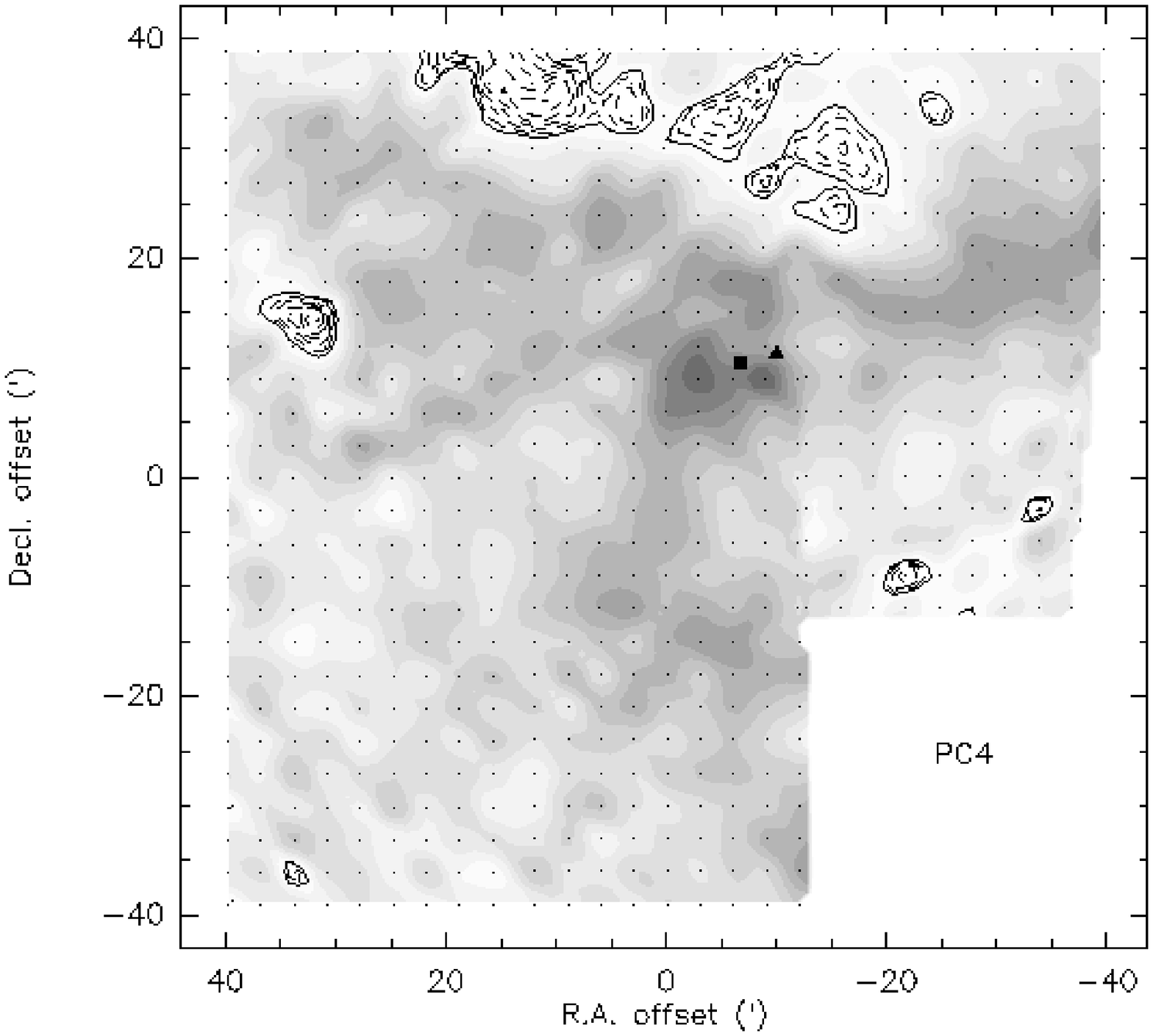,width=9cm,height=5.8cm,angle=270}
\caption[]{Eigenimages of the four first principal components calculated from
the $^{12}$CO(1--0) data. Negative contours are dashed (contour spacing 0.1),
greyscale represents positive values (for PC1 contour spacing is 0.05 else it is 0.2)
and the full line is the zero level isocontour
}
\label{fig:eigenimages}
\end{figure}

\begin{figure}
\psfig{file=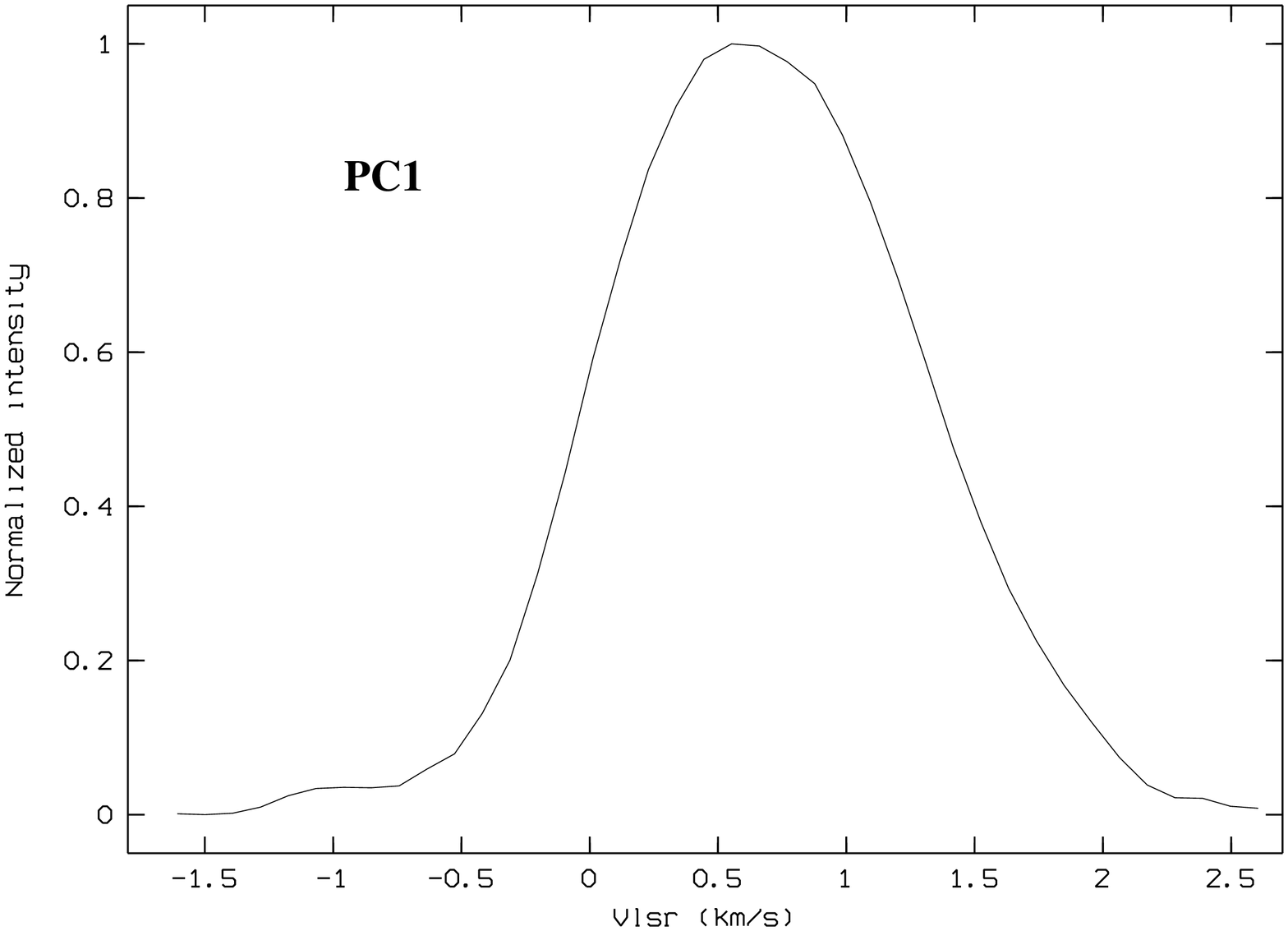,width=8cm,height=5.8cm,angle=270}
\psfig{file=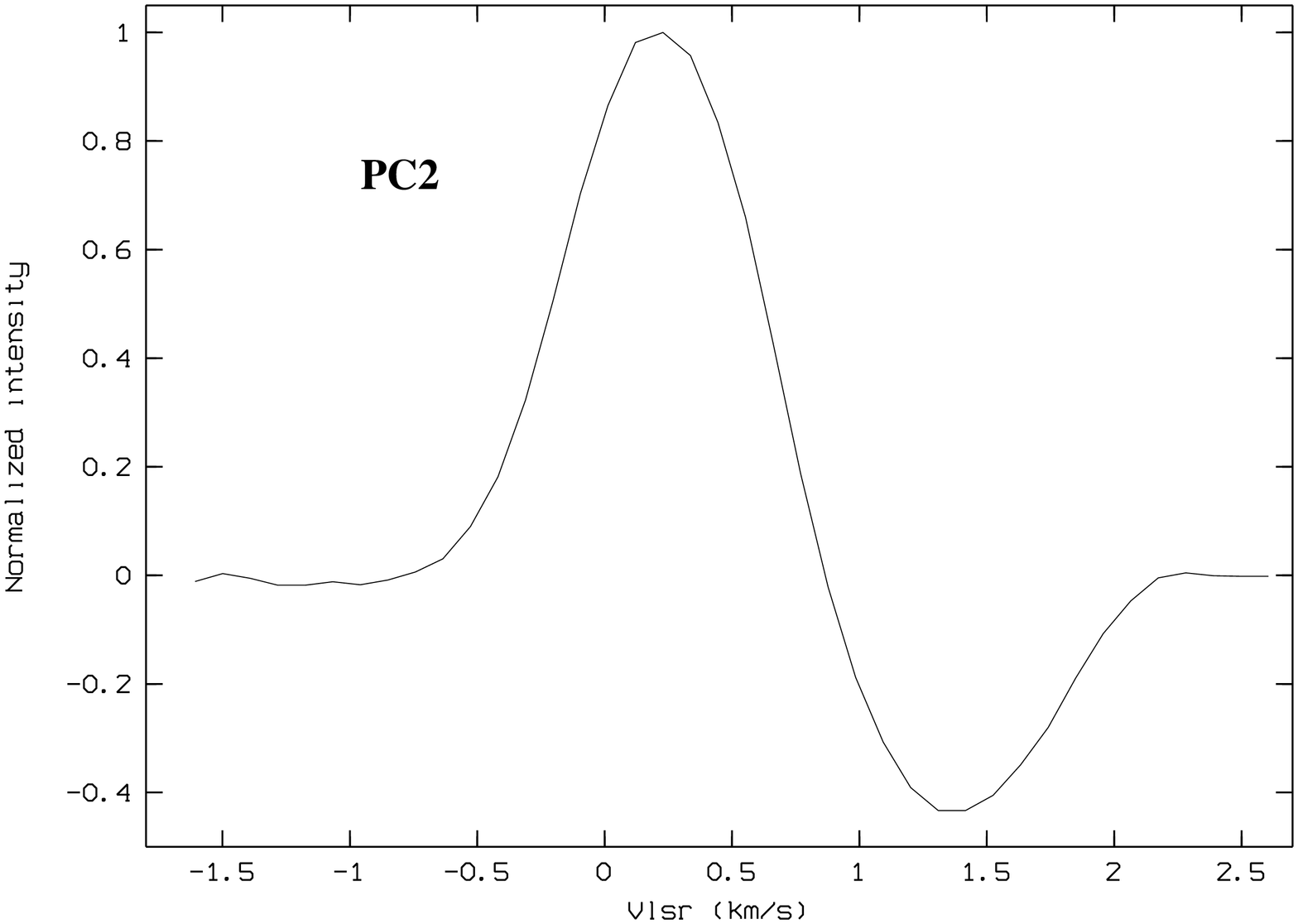,width=8cm,height=5.8cm,angle=270}
\psfig{file=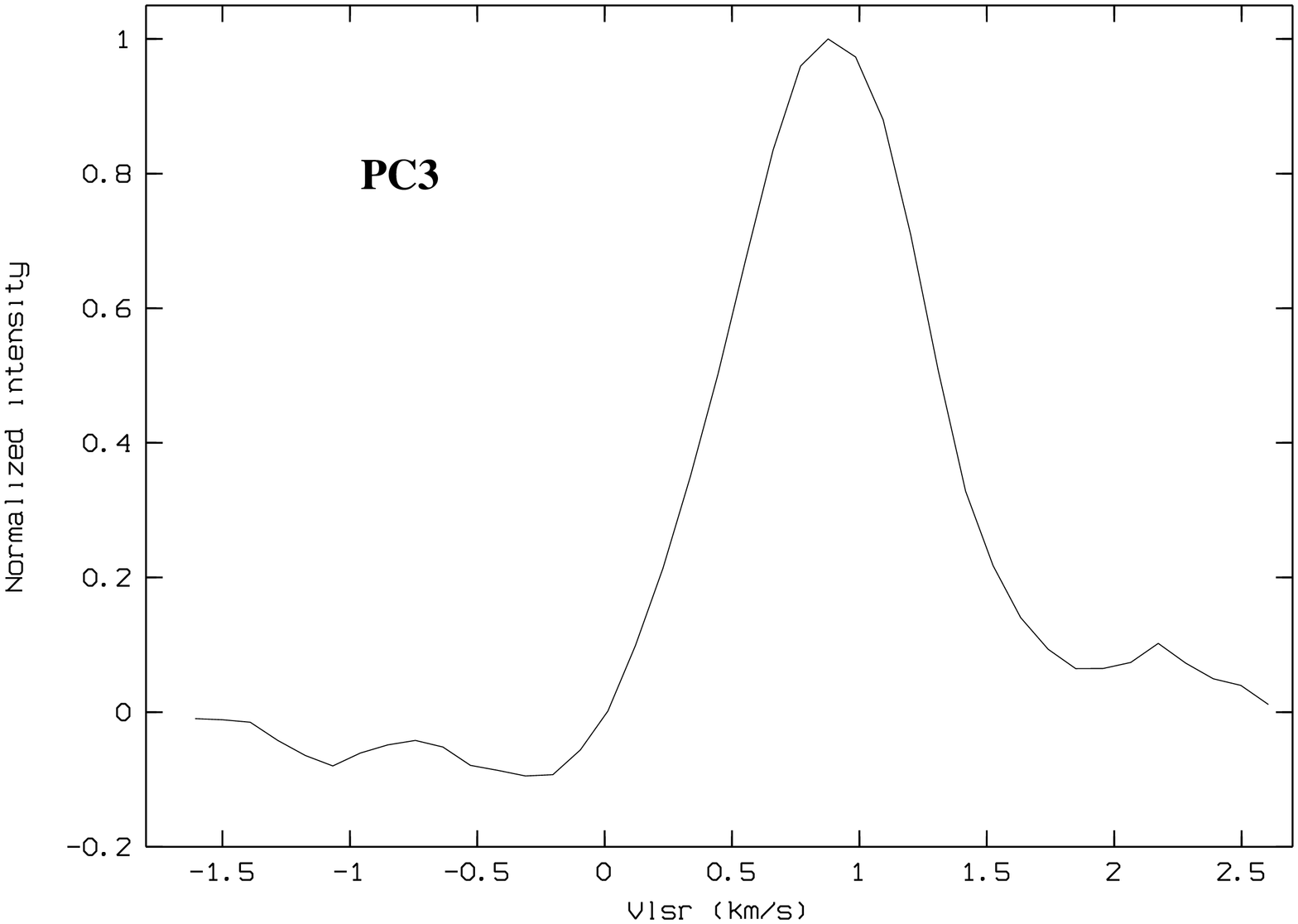,width=8cm,height=5.8cm,angle=270}
\psfig{file=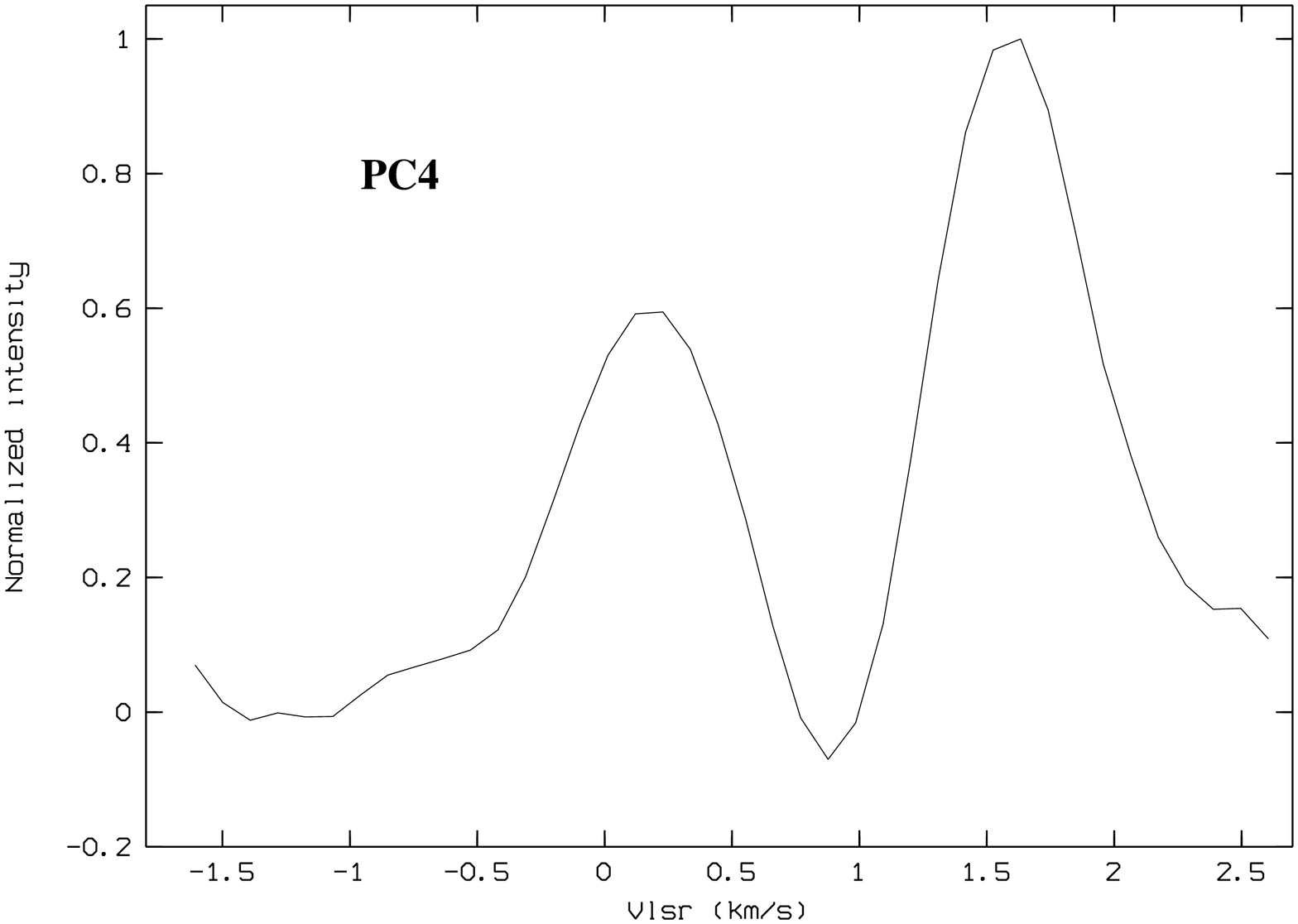,width=8cm,height=5.8cm,angle=270}
\caption[]{Eigenvectors used to generate the eigenimages
shown in Fig.~\ref{fig:eigenimages}}
\label{fig:eigenvectors}
\end{figure}

From $^{12}$CO(1--0) data a 648$\times$40 matrix was
constructed containing 648 observed spectra with 40 velocity channels. The
velocity range is [-1.6,+2.6]\,km\,s$^{-1}$ with a channel width of
0.1\,km\,s$^{-1}$. Before PCA analysis, the data are standardised, i.e. each
spectrum is mean-centred and normalised with its standard deviation.

Figures \ref{fig:eigenimages} and \ref{fig:eigenvectors} show our results.
The principal components 1-4 account for 71\%, 13\%, 7\% and 3\% of
the total variance, respectively. Together, these represent 95\% of the total
variance in our data. The remaining components are mostly pure noise and are
discarded. The importance of the first component shows that the spectra are
not independent from each other. As expected, there is a rather continuous
variation of line morphology throughout the cloud.

The first principal component, PC1, represents the average spectral profile
and the associated map shows the clumpy distribution of the main cloud. The
emission corresponds to the structures seen in channel maps at
$\sim$0.5\,km\,s$^{-1}$. The first eigenvector is roughly gaussian, with
position $\sim$0.6\,km\,s$^{-1}$ and width 1.4 km ${\rm s}^{-1}$.  The
following PCs describe successive corrections to the first approximation given
by PC1. Looking at the profiles, PC2 suggests a shift in velocity and PC4 a
change in the linewidth. PC3 indicates either a small change in the radial
velocity or, more likely, the presence of some narrower spectra.

PC2 mainly indicates velocity variations inside central region of L1642 and
separates this from the northern part of the field. The eigenvector indicates
that these sub-structures are around 0.2\,km\,s$^{-1}$ while the northern part
has a mean velocity of 1.2\,km\,s$^{-1}$. PC3 indicates a region of emission
around (0$\arcmin$, 0$\arcmin$) and (-9$\arcmin$, +8.5$\arcmin$) with a velocity
around 0.8\,km\,s$^{-1}$.  Finally, PC4 shows substructures in the core at
velocity $\sim$0.1\,km\,s$^{-1}$ especially around (-2.5$\arcmin$,
+9$\arcmin$), (-8, +17$\arcmin$) and (-8.5$\arcmin$, +9$\arcmin$). The map of
PC4 resembles the channel maps between 1.5 and 1.8\,km\,s$^{-1}$. The northern
part appears patchy, but it is difficult to say if the structure is caused
by real clumps or is simply due to noise fluctuations.

The PCA analysis follows a hierarchical scheme: first, it shows the central
part of L1642 (0.6\,km\,s$^{-1}$), secondly, it discriminates this from the
northern part (1.2\,km\,s$^{-1}$). Next it reveals additional velocity
structure inside L1642 itself (0.2, 0.8\,km\,s$^{-1}$), and finally fainter
emission at more positive velocity in the extreme north
($\sim$1.5\,km\,s$^{-1}$).

\subsection{Comparison of the analysis methods} \label{sect:comparison}

Channel maps are the standard method for analysing kinematic information of
molecular line maps. The presentation is mostly limited by the noise. If the
channel width is kept constant, the noise will dominate in the line wings. If,
on the other hand, noise is suppressed by integration over larger velocity
intervals, one loses all information of smaller velocity structures.
Furthermore, the signal level varies significantly over the map, and the
selection of integration intervals and contour levels is necessarily a
compromise. However, one can visually recognise continuous regions of
emission even when noise is quite significant. Channel maps do not provide
direct information on the spectral profiles because each channel is treated
completely separately. This information is obtained indirectly by looking at
the maps of consecutive velocity channels.

Both PCA and PMF present the observations as a linear combination of basic
spectral components which are extracted from the data. The methods carry out a
global fit, and can therefore discern features that are too weak to be
detected in any individual spectrum but which are repeated in many
observations. When the intensities of the spectral components are determined
for a given spectrum, the fit is done using all the relevant channels. The
methods are therefore most useful in the case of spectra with low
signal-to-noise ratios. The spectral features cannot, of course, be separated
unless their relative intensities change over the map. PCA and PMF
decompositions compress the information contained in the observation matrix to
be presented with just a few components. The result should therefore be
easier to interpret. In the case of PCA, however, this is complicated by
negative values present both in the profiles and in the component maps.

In the case of our L1642 observations, the three methods agree in the main
features. Comparison of the different methods shows that the PMF
method is especially able to produce a presentation that is at the same
time compact and easy to interpret. For example, a PMF decomposition with
$r$=4 reproduces practically all of the structure visible in the very complex
channel maps of $^{12}$CO. Because of the lower signal-to-noise ratio, only
three components could be detected in the C$^{18}$O data. When the rank of the
factorization is lower than the rank of the observation matrix, the
presentation is not exact. However, if $r$ is sufficiently large, the fit
residuals mostly consist of noise. The study of residuals of the PCA or PMF
fits is a very powerful tool for finding deviating spectra (see
Sect~\ref{sect:PMF}). Residuals identify abnormal spectra and show how their
profiles differ from the average behaviour.

In channel maps, one has the tendency to interpret intensity variations of
separate components as continuous velocity gradients. PMF and PCA 
explicitly assume that the emission is the sum of components at fixed radial
velocities. Velocity gradients are approximated as the sum of two or more
spectral components at different velocities, and with gradual changes in the
weights of the components. In the northern part of L1642, the channel maps of
$^{12}$CO $J$=1--0 and $J$=2--1 suggest a small velocity gradient towards the
north. PMF decomposition with $r$=4 presents the northern part with just one
velocity component. With $r$=5, the emission is divided into two components,
suggesting the presence of a velocity gradient. With higher $r$, the emission
is not divided any further. Within the noise limits the observations are,
therefore, consistent with {\em both} a small velocity gradient and the
presence of two separate emission components.

\section{Three-dimensional radiative transfer model} \label{appendix}

\subsection{Mass estimates} \label{appendix:3d}

The $^{13}$CO $J$=1--0 and $J$=2--1 observations of L1642 were modelled with a
fully three-dimensional model that consisted of 46$^3$ cells arranged on a
Cartesian grid. Along two dimensions, the column density distribution follows
the distribution obtained from the LTE calculations, and this fixes the size
of the model. Along the third dimension (i.e. along the line-of-sight), the
density distribution is assumed to be gaussian, with the extent of the cloud
similar along the line-of-sight as in the plane of the sky. In the model, the
linewidths are caused in roughly equal parts by microturbulence within cells
and macroscopic turbulence between cells. The mean velocity of cells on each
line-of-sight was shifted according to the velocity of the observed
$^{13}$CO(1--0) line. This enables direct comparison with the observed
spectra, but also creates velocity gradients that have a small effect on the
excitation. The density distribution was modified according to a volume
filling factor, $f$, by reducing the density by two orders of magnitude in a
randomly selected fraction 1-$f$ of the cells. This is only a crude
approximation of the true cloud structure, but such a model has already
been shown to reproduce some of the effects caused by density inhomogeneities (see e.g.
Park, Hong, \& Minh
\cite{park96}; Juvela \cite{juvela97}). The three-dimensional model is
fundamentally different from the one-dimensional models discussed in
Sect.~\ref{sect:colden}, and the comparison of the results should give a good
idea of the overall uncertainty of the mass estimates.

The parameter $f$ was varied between 0.1 and 1.0, and model calculations were
carried out for $T_{\rm kin}$=8-14\,K. The best fit with observations was
obtained with $T_{\rm kin}$=8\,K, $f=$0.4 and peak density of 1.7$\times
10^{4}$\,cm$^{-3}$. The total mass of the model cloud was 70\,$M_{\sun}$, 
very similar to the values obtained with spherically symmetric models. The
$\chi^2$-value of the fit was less than 10\% higher at 12\,K suggesting that the
temperature is not well constrained. Since the cloud size was fixed, the mass
decreases when the assumed temperature increases: for $T_{\rm kin}$=14\,K it
is 37\,$M_{\sun}$. The volume filling factor of the best fitting models
increases with temperature, and was already 1.0 at 14\,K. The $f$-parameter had
surprisingly little effect on the results, particularly at higher
temperatures. However, because of the macroturbulence, all models contained
equal amount of inhomogeneity in the velocity space.

\subsection{Comparison of true and LTE column densities}
\label{appendix:lte}

\begin{figure}
\resizebox{7.8cm}{!}{\includegraphics{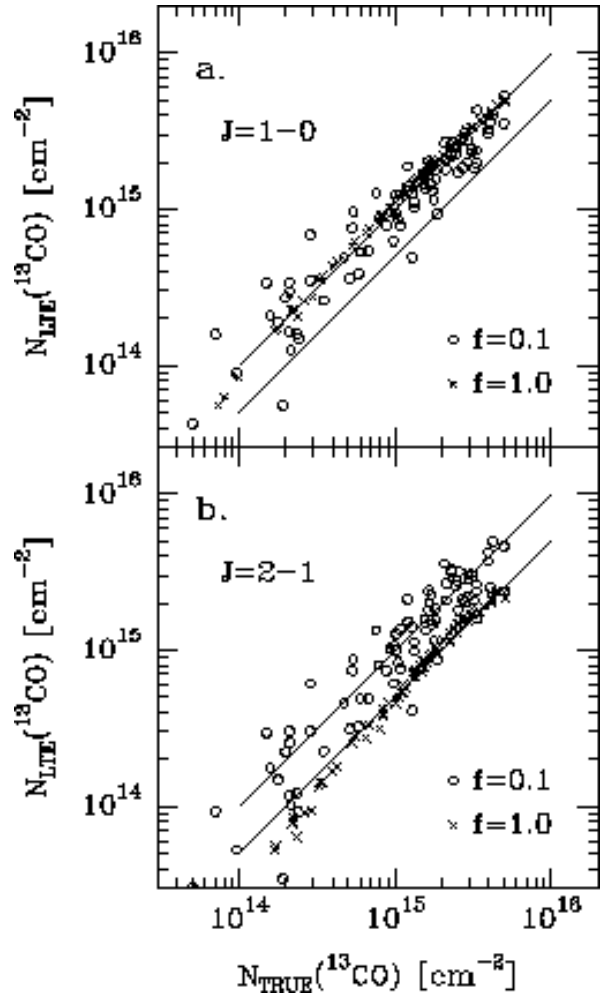}}
\caption[]{
Comparison of LTE column density estimates and the true, beam averaged column
densities in three-dimensional model clouds with $T_{\rm kin}$=10\,K and
volume filling factors 1.0 and 0.1. The lines indicate relations 1.0$\times
N_{\rm TRUE}$ and 0.5$\times N_{\rm TRUE}$ }
\label{fig:lte_fig}
\end{figure}

In order to further investigate the errors introduced by the LTE assumption, we
computed LTE column density estimates based on 3D cloud models with $T_{\rm
kin}$=10\,K and volume filling factors $f$=1.0 and $f$=0.1. The model
parameters were adjusted so that the model approximately reproduced the
$^{13}$CO observations of L1642. Spectra were calculated for the
$^{12}$CO(1--0) and $^{13}$CO(1--0) lines, and LTE analysis of the line ratios
of the calculated lines resulted in average excitation temperatures of $T_{\rm
ex}=$8.9\,K for the model with $f$=1.0 and $T_{\rm ex}=$8.4\,K for the model
with $f$=0.1. LTE column densities were computed based on these average
excitation temperatures and the calculated $^{13}$CO(1--0) spectra. In
Fig.~\ref{fig:lte_fig}, the LTE column densities are plotted against the
known, true column densities in the models.

For column densities derived from the $^{13}$CO(1--0) line, the LTE
calculations are quite accurate. In the case of $f$=0.1, the scatter is larger
because of the inhomogeneity of the cloud, and at high column densities the
LTE-estimates may be slightly biased towards low values. The LTE estimate is
not very sensitive to the assumed excitation temperature, and even a change of
2\,K in the value of $T_{\rm ex}$ has little effect on the derived values. The
volume-averaged, true excitation temperatures obtained from the radiative
transfer models are close to 5\,K, i.e. clearly below the LTE estimates of
8--9\,K. However, the large difference is mostly caused by low-density gas at
the edges of the cloud, and the bulk of emission comes from denser regions
which are much closer to thermalization.

We also plot in Fig.~\ref{fig:lte_fig} LTE estimates of the column density
derived from the model-calculated $J=$2--1 lines, assuming excitation
temperatures that are 1\,K lower than for the $J=$1--0 line. The column
density values are seen to be too low by up to a factor of two. In the case of
$f$=0.1, the density of the remaining clumps is higher, and since the gas is
closer to thermalization, the LTE estimates are also closer to the true column
density values. The results are more sensitive to the value of $T_{\rm ex}$
than in the case of the $J=$1--0 line. The column density values increase if
one assumes a lower excitation temperature. However, the LTE estimates of the
excitation temperature are {\em less} than 1\,K lower than for the
$J=$1--0 transition, and the same applies to the average excitation
temperature in the numerical models. The results show that the column density
estimates based on the $J=$2--1 transition are inherently more uncertain than
those derived from the $J=$1--0 observations.

\begin{thebibliography}{}

\bibitem[2000]{abraham00}
\'Abrah\'am P., Bal\'azs L.G., Kun M. 2000, A\&A 354, 645
\bibitem[2000]{bacmann00}
Bacmann A., Andr\'e P., Puget J.L., et al. 2000, A\&A 361, 555
\bibitem[1983]{blitz83}
Blitz L., Fich M. 1983,
in Kinematics, dynamics and structure of the Milky Way,
Dordrecht, D. Reidel Publishing Co., page 143.
\bibitem[1989]{booth89}
Booth R.S., Delgado G., Hagstr\"om, M., et al. 1989, A\&A 216, 315
\bibitem[1995]{brown95}
Brown A.G. Hartmann D., Burton W.B. 1995, A\&A 300,903
\bibitem[1992]{falgarone92}
Falgarone E., Puget J.-L., Perault M. 1992, A\&A 257, 715
\bibitem[1991]{falgarone91}
Falgarone E., Phillips T.G., Walker C.K. 1991, A\&A 378, 186
\bibitem[1988]{falgarone88}
Falgarone E., Puget J.-L., 1988, in Galactic and Extragalactic Star Formation,
eds. R.E. Pudritz, M. Fich, Kluwer, Dordrecht, P. 195
\bibitem[1988]{guelin88}
Guelin M., Cernicharo J., 1988,
in Molecular Clouds in the Milky Way and external galaxies,
Springer-Verlag
\bibitem[2000]{hearty00}
Hearty T., Fern\'andez M., Alcal\'a J.M., Covino E., Neuh\"auser R.
2000, A\&A 357, 681
\bibitem[1999]{heiles99}
Heiles C., Haffner M., Reynolds R., 1999,
ASP conf. Series 168, 211
\bibitem[1997]{heyer97}
Heyer M.H., Schloerb F.P. 1997, ApJ 475, 173
\bibitem[2000]{ingalls00}
Ingalls J., Bania T., Lane A., et al., 2000,
ApJ 535, 211
\bibitem[1997]{juvela97}
Juvela M. 1997, A\&A 322, 943
\bibitem[1996]{juvela96}
Juvela M., Lehtinen K., Paatero P. 1996, MNRAS 280, 616
\bibitem[1997]{kuntz97}
Kuntz K.D., Snowden S.L., Verter F. 1997,
ApJ 484, 245
\bibitem[1989]{langer89}
Langer W.D., Wilson R.W., Goldsmith P.F., Beichman C.A.,
1989, ApJ 337, 355
\bibitem[1987]{laureijs87}
Laureijs R.J., Mattila K., Schnur G. 1987,A\&A 184, 269
\bibitem[1988]{liljestrom88}
Liljestr\"om T., Mattila K. 1988, A\&A 196, 243
\bibitem[1989]{liljestrom89}
Liljestr\"om T., Mattila K., Friberg P. 1989, A\&A 210, 337
\bibitem[1991]{liljestrom91}
Liljestr\"om T. 1991, A\&A 244,483
\bibitem[1987]{MH87}
Murtagh F., Heck A., 1987, Multivariate Data Analysis. 
Ap. Sp. Sc. Lib., Dordrecht: Reidel, p. 13 
\bibitem[1999]{paatero99}
Paatero P. 1999, Journal of Comp. and Graph. Stat., vol. 8, 854
\bibitem[2000]{padoan00}
Padoan P., Juvela M., Bally J., Nordlund \AA 2000, ApJ 529, 259
\bibitem[1996]{park96}
Park Y.S., Hong S.S., Minh Y.C. 1996, A\&A 312, 981 
\bibitem[1990]{reipurth90}
Reipurth B., Heathcote S. 1990, A\&A 229, 527
\bibitem[1979]{reynolds79}
Reynolds R.J., Odgen P.M. 1979, ApJ 229, 942
\bibitem[1987]{sandell87}
Sandell G., Reipurth B., Gahm G. 1987, A\&A 181, 283
\bibitem[1999]{sfeir99}
Sfeir D.M., Lallement R., Crifo F., Welsh B.Y. 1999,
A\&A 346, 785
\bibitem[1982]{taylor82}
Taylor M., Taylor K., Vaile R., 1982,
Proc. ASA 4, 440
\bibitem[1976]{ulich76}
Ulich B.L., Haas R.W. 1976, ApJS 30, 247
\bibitem[1997]{ungerechts97}
Ungerechts H.,  Bergin E.A., Goldsmith P.F., et al.
1997, ApJ 482, 245
\bibitem[1994]{ward94}
Ward-Thompson D., Scoot P.F., Hills R.E., Andr\'e P. 1994, MNRAS 268, 276
\bibitem[1999]{ward99}
Ward-Thompson D., Motte F., Andr\'e P. 1999, MNRAS 305, 143
\end{thebibliography}
\end{document}